\documentclass[11pt,a4paper]{article}
\usepackage{graphics}
\usepackage{hyperref}
\usepackage{graphicx,epsfig,epstopdf}
\usepackage{bm}
\usepackage{amssymb,amsmath,amsfonts,latexsym}
\usepackage{dcolumn}
\usepackage[italic]{hepnames}
\usepackage[utf8]{inputenc}
\usepackage[T1]{fontenc}
\usepackage{microtype}
\usepackage[margin=1in]{geometry}
\usepackage{caption}
\usepackage{subcaption}
\usepackage{tikz}
\usepackage{pgfplots}
\pgfplotsset{compat=1.18}
\usepackage{float}
\usepackage{multirow}
\usepackage{breqn}
\usepackage{booktabs,longtable,array}
\usepackage{enumitem}
\usepackage[T1]{fontenc}
\usepackage[utf8]{inputenc}
\usepackage{lmodern}

\hypersetup{colorlinks=true,linkcolor=blue,citecolor=blue,urlcolor=blue}
\makeatletter
\let\cat@comma@active\@empty
\makeatother

\providecommand{\Bc}{B_c}
\providecommand{\etac}{\eta_c}
\providecommand{\psij}{J/\psi}

\providecommand{\psione}{\psi_1}
\providecommand{\psitwo}{\psi_2}
\providecommand{\psithree}{\psi_3}
\providecommand{\etatwo}{\eta_{c2}}

\begin{document}

\title{A unified study of the \(B_c\) meson: from spectrum and form factors to weak and radiative decays}

\author{ Vikas Patel$^{1}$, Chetan Lodha$^{2}$, Raghav Chaturvedi$^{2}$, A. K. Rai$^{2}$\\[0.4em]
\small $^{1}$Department of Physics, Uka Tarsadia University, Bardoli 394250, Gujarat, India\\
\small $^{2}$Department of Physics, Sardar Vallabhbhai National Institute of Technology, Surat-395007, Gujarat, India\\
\small \texttt{iamchetanlodha@gmail.com}}

\date{}

\maketitle

\begin{abstract}
The $B_c$ meson constitutes a unique system for investigating heavy-hadron dynamics, since it exhibits quarkonium-like bound-state structure while decaying predominantly through weak interactions. In this work, we present a unified study of $B_c$-meson spectroscopy, decay constants, weak decays, radiative transitions, and Regge trajectories within a single framework. The mass spectrum is computed in a screened potential model including relativistic kinetic-energy corrections and spin-dependent interactions, from which we obtain both spin-averaged and spin-resolved states together with the corresponding pseudoscalar and vector decay constants. Using these spectroscopic inputs, we then analyze weak decays within a mass-updated three-point QCD sum-rule framework for transitions to $S$-, $P$-, and $D$-wave charmonium states, as well as to final states containing charmonium and $D^{(*)}_{(s)}$ mesons. In this part of the analysis, the hadronic thresholds, Borel-window prescriptions, Lorentz decompositions, and decay-width expressions of the underlying sum-rule formulations are retained, while the heavy-quark masses are updated to $m_c=1.48~\mathrm{GeV}$ and $m_b=4.90~\mathrm{GeV}$, leading to a controlled refit of the overall normalization rather than a full rederivation of all perturbative and condensate contributions. We further investigate purely leptonic decay widths and radiative $E1$ and $M1$ transitions, and examine the Regge behavior of the resulting spectrum. The present study therefore provides a coherent description of the $B_c$ meson in which the spectroscopic wave functions determine the short-distance couplings that enter both weak and electromagnetic observables, while the Regge analysis serves as a complementary global consistency test of the same dynamical picture.
\end{abstract}

\section{Introduction}

The $B_c$ meson occupies a special position in hadron physics as the only conventional meson composed of two heavy quarks with different flavour content, $\bar bc$ or $b\bar c$. Because of this unequal heavy-quark composition, it combines features of heavy quarkonia and open-flavour heavy hadrons in a single system. On the one hand, its dynamics are governed by a compact heavy-quark bound state and are therefore sensitive to interquark forces, spin-dependent interactions, and radial-orbital level structure. On the other hand, unlike charmonium and bottomonium, the $B_c$ family does not admit strong annihilation into gluons or light hadrons in the same way, and the ground state decays only weakly. This makes the $B_c$ sector an especially useful laboratory for studying how spectroscopy, short-distance wave functions, weak transition amplitudes, and electromagnetic observables are linked within one heavy-meson system \cite{KwongRosner1991,EichtenQuigg1994,Gershtein1995PRD,GuptaJohnson1996,BrambillaVairo2000,Godfrey2004,Ebert2011Regge,Li2019NRQM,Patnaik2024Review}.

The theoretical importance of the $B_c$ meson was recognized well before its experimental discovery. Early quark-model and perturbative studies showed that the lowest $B_c$ states should lie below the open-flavour thresholds for strong decay and should therefore be relatively narrow, with widths controlled primarily by weak interactions \cite{KwongRosner1991,EichtenQuigg1994,Gershtein1995PRD,ChangChen1992Prod}. This immediately distinguished the system from ordinary heavy quarkonia and suggested that it could serve as a bridge between two traditionally separate areas of heavy-hadron phenomenology: the spectroscopy of heavy bound states and the weak decay of flavour-carrying hadrons. In the decades that followed, the $B_c$ family became a testing ground for constituent-quark models, relativistic potential models, Bethe--Salpeter approaches, QCD sum rules, lattice QCD, and more recent screened-potential and coupled-channel descriptions \cite{Patnaik2024Review,ChangChen1994Decays,Gershtein1995PhysUsp,Kiselev2000,Ebert2003Prop,AbdElHadyMunozVary2000,IvanovKornerSantorelli2006,Tang2018,Harrison2020Form,Colangelo2022Relations}.

From the spectroscopic point of view, the $B_c$ system is richer than equal-mass heavy quarkonia because the unequal quark masses break the simple symmetry between the two constituents. As a result, the pattern of spin splittings, radial excitations, orbital multiplets, and mixing effects carries information about both the short-distance Coulombic part of the interaction and the confining part of the potential. This makes the $B_c$ meson particularly suitable for testing whether a given model can describe not only the gross mass scale of the ground state, but also the resolved pattern of excited states and fine structure across the $S$-, $P$-, $D$-, and higher-wave sectors \cite{EichtenQuigg1994,Gershtein1995PRD,Godfrey2004,Ebert2011Regge,Li2019NRQM,Ebert2003Prop,Ortega2020,Li2023Survey}. In a broader meson-spectroscopy context, recent studies of strangeonium, kaon, $D$, and charm-strange systems within Regge and related phenomenological frameworks have further emphasized the continuing value of global trajectory analyses and systematic state classification in understanding hadron spectra \cite{Oudichhya:2023KaonStrangeonium,Oudichhya:2024DMeson,Oudichhya:2024CharmStrangePoS,Oudichhya:2025StrangeoniumDAE,Oudichhya:2026KaonDAE,Oudichhya:2026CharmStrangeSpringer}. Closely related screened-potential studies of the $D$ meson, charm-strange mesons, and the $B$ and $B_s$ mesons provide useful benchmarks for the potential dynamics, relativistic corrections, decay constants, radiative widths, and Regge organization employed in the present work \cite{Patel:2021Dmeson,Patel:2021CharmStrange,Patel:2024BBsScreened}. Likewise, recent studies of strangeonium, kaonic, charm-bound, fully heavy, and beauty-flavoured systems, including works that analyze meson--tetraquark interplay, illustrate how modern spectroscopy increasingly attempts to connect conventional and exotic configurations within a unified phenomenological setting \cite{Lodha:2024Strangeonium,Lodha:2025Dmesons,Lodha:2025Kaons,Lodha:2024AllLight,Lodha:2025LightStrange,Patel:2025rsf,Patel:2025bmc,Patel:2025bdu,Lodha:2026iww}. At the same time, the wave functions extracted from the spectroscopic calculation determine short-distance observables such as decay constants and weak-annihilation amplitudes, so the spectroscopy is not merely a static input but the structural basis for the entire phenomenology.

Experimentally, the $B_c$ programme has evolved from discovery to precision measurement. After the first observation by the CDF Collaboration \cite{CDF1998PRL,CDF1998PRD}, subsequent measurements at the Tevatron and the LHC significantly improved the knowledge of the ground-state mass, production properties, and lifetime \cite{CDF2006Lifetime,D02008Lifetime,LHCb2012ProdMass,Aaij2014BcLifetime,LHCb2020Mass}. The observation of excited structures by ATLAS, CMS, and LHCb further broadened the phenomenological landscape and provided valuable experimental anchors beyond the ground state \cite{ATLAS2014Excited,CMS2019Excited,LHCb2019Excited}. In parallel, the Particle Data Group averages now supply standard reference values against which modern theoretical calculations can be calibrated \cite{PDG2024BflavoredHadrons}. The combination of a well-measured ground state and emerging information on excited states makes the current period especially suitable for a consolidated $B_c$ analysis.

The weak-decay phenomenology of the $B_c$ meson is equally distinctive. Because the ground state lies below the lowest open-flavour strong-decay threshold, its total width receives contributions from three mechanisms: the decay of the $b$ quark with the charm quark as spectator, the decay of the $c$ quark with the bottom quark as spectator, and the weak annihilation of the $\bar bc$ pair \cite{ChangChen1994Decays,Gershtein1995PhysUsp,Kiselev2000,LiuChao1997}. This three-component structure gives the $B_c$ meson a richer weak-decay pattern than ordinary quarkonia and makes it particularly sensitive to decay constants, semileptonic form factors, nonleptonic amplitudes, helicity structure, and lepton-flavour-universality ratios. In recent years, exclusive channels such as $B_c\to J/\psi \ell\bar\nu_\ell$, $B_c\to\eta_c \ell\bar\nu_\ell$, $B_c\to \chi_{cJ}\ell\bar\nu_\ell$, and nonleptonic modes involving charmonium and open-charm mesons have become central to the phenomenology of the system \cite{ColangeloDeFazio1999Rad,ColangeloDeFazio2000HQSS,Ivanov2000Semileptonic,Dutta2017TauNu,Harrison2020RJpsi}.

A wide range of theoretical tools has been developed to address these observables. Potential and constituent-quark models established the basic spectroscopic structure of the low-lying states \cite{EichtenQuigg1994,Gershtein1995PRD,Godfrey2004,RaiVinodkumar2006}. Relativistic quark models and Bethe--Salpeter treatments extended the discussion to decay constants, form factors, and radiative transitions \cite{AbdElHadyMunozVary2000,Tang2018,AbdElHadyLodhiVary1999,Ebert2003BcCharmD,Ebert2003BsB,AbdElHadySpenceVary2005,Issadykov2018,Yao2021}. QCD sum rules and lattice QCD have supplied increasingly important benchmarks for masses, decay constants, and semileptonic form factors, especially in the phenomenologically important $B_c\to J/\psi$ and $B_c\to\eta_c$ channels \cite{Harrison2020Form,Harrison2020RJpsi,Davies1996,Allison2005,Gregory2010,Dowdall2012,Baker2014,Feng2023ThreeLoop}. More recently, screened-potential models, Regge analyses, and broad surveys of the upper spectrum have sharpened the discussion of how higher excitations and trajectory patterns should be interpreted in the $B_c$ sector \cite{Patnaik2024Review,Ortega2020,Li2023Survey,RadfordRepko2007,DevlaniRai2014,Monteiro2017,Akbar2019}. Related developments in meson phenomenology, including Regge analyses of kaon, strangeonium, $D$, and charm-strange states and relativistic flux-tube studies of heavy-light mesons, further demonstrate the continuing usefulness of global spectroscopic frameworks for organizing hadronic excitations across different flavour sectors \cite{Oudichhya:2023KaonStrangeonium,Oudichhya:2024DMeson,Oudichhya:2024CharmStrangePoS,Jakhad:2025BottomMesons,Jakhad:2026CharmedMesons,Purohit:2022LIMYP}.

Motivated by this broader context, the present work develops a unified phenomenological analysis of the $B_c$ meson in which spectroscopy provides the starting point for all subsequent observables. We employ a screened-potential framework that preserves the short-distance Coulombic behaviour of one-gluon exchange while softening the long-distance confining interaction. This choice is advantageous for the $B_c$ family because it retains the successful low-lying heavy-quarkonium description while allowing a more realistic treatment of the higher radial and orbital excitations. Relativistic corrections to the kinetic-energy operator and perturbative spin-dependent interactions are included consistently, so that the same framework can be used to obtain spin-averaged masses, resolved multiplets, short-distance wave functions, and overlap-sensitive observables.

On this basis we investigate the spectroscopy of the $B_c$ family, the pseudoscalar and vector decay constants, purely leptonic and weak-annihilation observables, radiative transitions, and Regge trajectories. We also connect the spectroscopic picture to weak-decay phenomenology through a QCD sum-rule treatment of transitions to $S$-, $P$-, and $D$-wave charmonia and to nonleptonic channels involving charmonium plus $D^{(*)}_{(s)}$ mesons, for which recent dedicated analyses provide useful benchmarks \cite{Wu:2024gcq,Lu:2025bvi,Wu:2024dzn,Lu:2025usr}. The goal is not merely to present isolated tables of results, but to show how the mass spectrum, short-distance couplings, weak amplitudes, electromagnetic widths, and global trajectory structure emerge as mutually related consequences of a single model.

The article is organized as follows. In Sec.~2 we present the theoretical framework, including the screened-potential description, the spin-dependent interaction, the treatment of decay constants, the form-factor decomposition of the weak matrix elements, the QCD sum-rule construction of the transition form factors, and the formalism for purely leptonic, radiative, and Regge analyses. In Sec.~3 we specify the QCD sum-rule inputs and the numerical implementation. Section~4 contains the full set of results and comparisons with representative theoretical studies, with all physical interpretation consolidated there. Finally, Sec.~5 summarizes the main conclusions. In this way, the manuscript is intended to read as a coherent $B_c$-focused original study, informed by the literature but organized around the internal consistency of the present phenomenological framework.

\section{Theoretical framework}

\subsection{Screened-potential description of the $B_c$ system}

In the present work, the $B_c$ meson is treated as a heavy quark--antiquark bound state in a semi-relativistic constituent-quark framework. Such an approach is well suited to the $B_c$ family because the system is compact enough for short-distance one-gluon-exchange dynamics to remain important, while its higher radial and orbital excitations are still sensitive to the long-distance behaviour of confinement \cite{EichtenQuigg1994,Gershtein1995PRD,Godfrey2004,Ebert2011Regge,Li2019NRQM,Patnaik2024Review,Ortega2020}. The unequal masses of the charm and bottom constituents make the $B_c$ sector especially valuable, since the spectroscopy reflects the interplay of two distinct heavy-quark scales rather than a single nearly symmetric mass scale as in charmonium or bottomonium.

The effective Hamiltonian is written as
\begin{equation}
	H=\sqrt{\mathbf{p}^{\,2}+m_c^2}+\sqrt{\mathbf{p}^{\,2}+m_b^2}+V(r),
	\label{eq:Hamiltonian_main}
\end{equation}
where $\mathbf{p}$ is the relative momentum of the quark--antiquark pair, $m_c$ and $m_b$ are the constituent masses of the charm and bottom quarks, and $V(r)$ denotes the effective interaction potential. The rest-mass contribution $m_c+m_b$ is separated from the square-root terms, and the remaining kinetic-energy operator is expanded in powers of $\mathbf{p}^2$,
\begin{eqnarray}
	K.E.\equiv T_{\mathrm{rel}} &=& \frac{\mathbf{p}^{2}}{2}\left(\frac{1}{m_c}+\frac{1}{m_b}\right)
	-\frac{\mathbf{p}^{4}}{8}\left(\frac{1}{m_c^{3}}+\frac{1}{m_b^{3}}\right)
	+\frac{\mathbf{p}^{6}}{16}\left(\frac{1}{m_c^{5}}+\frac{1}{m_b^{5}}\right)\nonumber\\
	&&-\frac{5\mathbf{p}^{8}}{128}\left(\frac{1}{m_c^{7}}+\frac{1}{m_b^{7}}\right)
	+\frac{7\mathbf{p}^{10}}{256}\left(\frac{1}{m_c^{9}}+\frac{1}{m_b^{9}}\right).
	\label{eq:KE_main}
\end{eqnarray}
Here $K.E.$ denotes the relative kinetic energy after subtracting the constituent rest masses. The relevant expansion variables are $\langle\mathbf{p}^{2}\rangle/m_c^2$ and $\langle\mathbf{p}^{2}\rangle/m_b^2$; the scale of $\mathbf{p}^2$ is therefore not introduced as an external parameter but is fixed by the optimized variational wave function for each state, being of order the squared Gaussian width. With the widths obtained below, this momentum scale remains well below $m_b^2$ and is moderate compared with $m_c^2$, so the expansion is used as a semi-relativistic approximation with terms retained through $\mathbf{p}^{10}$. Such expansions are widely used in phenomenological studies of heavy-heavy systems \cite{Ebert2003Prop,RaiVinodkumar2006,DevlaniRai2014} and are also consistent with related screened-potential applications to open-charm, charm-strange, and bottom mesons \cite{Patel:2021Dmeson,Patel:2021CharmStrange,Patel:2024BBsScreened}.

The interquark interaction is decomposed into a leading central term and a subleading correction,
\begin{equation}
	V(r)=V^{(0)}(r)+\left(\frac{1}{m_c}+\frac{1}{m_b}\right)V^{(1)}(r)+\mathcal{O}\!\left(\frac{1}{m^2}\right),
	\label{eq:potential_decomp_main}
\end{equation}
where
\begin{equation}
	V^{(0)}(r)=V_V(r)+V_S(r).
\end{equation}
The vector component is taken as the Coulombic one-gluon-exchange interaction,
\begin{equation}
	V_V(r)=-\frac{4}{3}\frac{\alpha_s}{r},
\end{equation}
while the scalar component is modeled through a screened confining term,
\begin{equation}
	V_S(r)=\frac{A}{\xi_s}\left(1-e^{-\xi_s r}\right)+V_0.
\end{equation}
Here $\alpha_s$ is the strong coupling, $A$ controls the strength of confinement, $\xi_s$ is the screening parameter, and $V_0$ is an additive constant. The screening plays a central role in the present model: it preserves the approximately linear rise of the confining interaction at moderate distances while softening the potential at larger separation, which is especially relevant for higher excitations and for the global structure of the upper spectrum \cite{Ebert2011Regge,Ortega2020,Li2023Survey,Patel:2021Dmeson,Patel:2021CharmStrange,Patel:2024BBsScreened}.

Including the leading correction, the full central potential becomes
\begin{equation}
	V(r)= -\frac{4}{3}\frac{\alpha_s}{r}
	+\frac{A}{\xi_s}\left(1-e^{-\xi_s r}\right)
	+V_0
	+\left(\frac{1}{m_c}+\frac{1}{m_b}\right)V^{(1)}(r)
	+\mathcal{O}\!\left(\frac{1}{m^2}\right),
	\label{eq:full_potential_main}
\end{equation}
with
\begin{equation}
	V^{(1)}(r)=-\frac{C_FC_A\alpha_s^2}{4r^2}.
	\label{eq:V1_main}
\end{equation}
This correction term accounts for the leading short-distance improvement beyond the static central interaction.

The model parameters adopted in the present analysis are summarized in Table~\ref{Table:Parameter}. These inputs are kept fixed throughout the spectroscopic sector and provide the basis for all later observables derived from the wave functions.

\begin{table}[H]
	\caption{Model parameters adopted for the screened-potential description of the $B_c$ meson. Here \(\alpha_s\) is the fixed effective strong coupling used in the Coulombic and spin-dependent terms, \(\alpha_c=4\alpha_s/3\) is the corresponding color-Coulomb coefficient shown for convenience, and \(\xi_s\) denotes the screening parameter of the confining interaction. The Gaussian variational width is denoted separately by \(\beta\) in the state-by-state analysis.}
	\label{Table:Parameter}
	\centering
	\begin{tabular}{cccccccccc}
		\hline\noalign{\smallskip}
		\hline\noalign{\smallskip}
		Meson & $\alpha_s$ & $\alpha_c$ & $m_c$ & $m_b$ & $n_f$ & $\lambda$ & $\xi_s$ & $A$ & $V_0$ \\
		&&&(GeV)&(GeV)&&&&$(\mathrm{GeV/fm})$&\\
		\hline
		$B_c$ & 0.319 & 0.425 & 1.55 & 4.88 & 5 & 0.195 & 0.04 & 0.200 & -0.271 \\
		\hline\noalign{\smallskip}
	\end{tabular}
\end{table}

The couplings in Table~\ref{Table:Parameter} are fixed phenomenological inputs for the $B_c$ calculation, not state-by-state running couplings. In particular, $\alpha_s$ sets the strength of the one-gluon-exchange part of the potential and $\alpha_c$ is not an additional independent parameter; it is the color factor $4/3$ multiplied by $\alpha_s$. We keep these values fixed to avoid introducing an extra scale ambiguity into the variational calculation, while the scale dependence of short-distance QCD is effectively absorbed into the fitted potential parameters.

To solve the radial problem, we employ Gaussian trial wave functions in both coordinate and momentum space,
\begin{equation}
	R_{nl}(\beta,r)=\beta^{3/2}
	\left[\frac{2(n-1)!}{\Gamma\!\left(n+l+\frac12\right)}\right]^{1/2}
	(\beta r)^l e^{-\beta^2 r^2/2}
	L_{n-1}^{\,l+1/2}(\beta^2 r^2),
	\label{eq:coord_wf_main}
\end{equation}
and
\begin{equation}
	R_{nl}(\beta,p)=\frac{(-1)^n}{\beta^{3/2}}
	\left[\frac{2(n-1)!}{\Gamma\!\left(n+l+\frac12\right)}\right]^{1/2}
	\left(\frac{p}{\beta}\right)^l
	e^{-p^2/(2\beta^2)}
	L_{n-1}^{\,l+1/2}\!\left(\frac{p^2}{\beta^2}\right),
	\label{eq:mom_wf_main}
\end{equation}
where $\beta$ is the variational width parameter and $L_{n-1}^{\,l+1/2}$ denotes the associated Laguerre polynomial. This basis is convenient because it yields stable analytic matrix elements for both the central and spin-dependent pieces of the Hamiltonian while remaining flexible enough to describe radial and orbital excitations \cite{RaiVinodkumar2006,DevlaniRai2014,Abreu2020}.

The variational parameter $\beta$ is fixed through the virial condition
\begin{equation}
	\left\langle\frac{\mathbf{p}^{2}}{2\mu}\right\rangle
	=
	\frac{1}{2}\left\langle r\frac{dV}{dr}\right\rangle,
	\qquad
	\mu=\frac{m_cm_b}{m_c+m_b},
	\label{eq:virial_main}
\end{equation}
which is applied separately to each radial and orbital state. Once the expectation values are determined, the spin-averaged mass is defined as
\begin{equation}
	M_{SA}=M_P+\frac{3}{4}(M_V-M_P),
	\label{eq:msa_main}
\end{equation}
where $M_P$ and $M_V$ denote, respectively, the pseudoscalar spin-singlet and vector spin-triplet masses belonging to the same radial $S$-wave level. For higher orbital multiplets we use the corresponding spin-weighted center-of-weight definition,

\begin{equation}
	M_{CW,n}=\frac{\sum_J(2J+1)m_{nJ}}{\sum_J(2J+1)}.
	\label{eq:mcw_main}
\end{equation}
These quantities provide the natural bridge between the central-potential spectrum and the inclusion of fine and hyperfine splittings.

\subsection{Spin-dependent interaction and resolved spectrum}

The spin-averaged masses obtained from the central Hamiltonian are resolved into physical states by means of the standard spin-dependent interaction,
\begin{eqnarray}
	V_{\mathrm{spin}}(r) &=&
	\left(\frac{\mathbf{L}\!\cdot\!\mathbf{S}_c}{2m_c^2}
	+\frac{\mathbf{L}\!\cdot\!\mathbf{S}_{\bar b}}{2m_b^2}\right)
	\left(-\frac{1}{r}\frac{dV^{(0)}(r)}{dr}
	+\frac{8}{3}\alpha_s\frac{1}{r^3}\right)\nonumber\\
	&&+\frac{4}{3}\alpha_s\frac{1}{m_cm_b}\frac{\mathbf{L}\!\cdot\!\mathbf{S}}{r^3}
	+\frac{4}{3}\alpha_s\frac{2}{3m_cm_b}\,
	(\mathbf{S}_c\!\cdot\!\mathbf{S}_{\bar b})\,4\pi\delta(\mathbf{r})\nonumber\\
	&&+\frac{4}{3}\alpha_s\frac{1}{m_cm_b}
	\left[3(\mathbf{S}_c\!\cdot\!\mathbf{n})(\mathbf{S}_{\bar b}\!\cdot\!\mathbf{n})
	-(\mathbf{S}_c\!\cdot\!\mathbf{S}_{\bar b})\right]\frac{1}{r^3},
	\qquad \mathbf{n}=\frac{\mathbf{r}}{r}.
	\label{eq:spin_interaction_main}
\end{eqnarray}
This interaction contains the spin-orbit, contact spin-spin, and tensor components responsible for fine and hyperfine splittings. In the $B_c$ system, these terms are especially important because the unequal quark masses break the simple equal-mass pattern familiar from charmonium and bottomonium, leading to a richer resolved structure across the $S$-, $P$-, and higher-wave sectors \cite{Godfrey2004,Li2019NRQM,Ebert2003Prop}.

For observables controlled by the short-distance wave function, we use the spin-corrected wave function at the origin,
\begin{equation}
	R_{nJ}(0)=R(0)\left[1+(SF)_J\frac{\langle \varepsilon_{SD}\rangle_{nJ}}{M_{SA}}\right],
	\label{eq:rnj_main}
\end{equation}
with the spin-averaged value
\begin{equation}
	R(0)=\frac{R_P+3R_V}{4}.
	\label{eq:r0avg_main}
\end{equation}
Here $R_P$ and $R_V$ denote the radial wave functions at the origin for the pseudoscalar spin-singlet and vector spin-triplet $S$-wave states belonging to the same radial level, respectively. The factor of three multiplying $R_V$ reflects the spin degeneracy of the vector state, so that $R(0)$ represents the spin-weighted average short-distance wave function used as the common input for the subsequent spin correction.
This prescription is useful because it allows the spectroscopic fine structure and the short-distance decay observables to remain tied to the same underlying bound-state framework rather than being treated independently.

\subsection{Decay constants}

The pseudoscalar and vector decay constants are obtained from the momentum-space wave functions through a Van--Royen--Weisskopf-type relation supplemented by relativistic corrections,
\begin{equation}
	\begin{aligned}
		f_{P,V}
		&=
		\left(\frac{12}{M_{P,V}}\right)^{1/2}
		\int \frac{d^3p}{(2\pi)^3}
		\left(\frac{E_c(p)+m_c}{2E_c(p)}\right)^{1/2}
		\left(\frac{E_{\bar b}(p)+m_{\bar b}}{2E_{\bar b}(p)}\right)^{1/2}
		\\
		&\qquad\times
		\left[
		1+\frac{\lambda_{P,V}\,p^2}
		{\big(E_c(p)+m_c\big)\big(E_{\bar b}(p)+m_{\bar b}\big)}
		\right]
		\Phi_{P,V}(p),
	\end{aligned}
	\label{eq:decayconstant_main}
\end{equation}
where $\lambda_P=-1$ and $\lambda_V=-1/3$. In this form the decay constants remain directly tied to the same short-distance wave functions that determine the hyperfine structure and weak-annihilation amplitudes. Since the decay constants are controlled predominantly by the behaviour of the wave function near the origin, they provide the most direct bridge between the spectroscopic part of the calculation and the weak-decay phenomenology discussed later \cite{Ebert2003Prop,Tang2018,Kiselev2004Leptonic,SunNiChen2023}.

\subsection{Transition form factors and hadronic matrix elements}

The weak-decay sector is formulated in terms of hadronic matrix elements of the charged current
\begin{equation}
	J_\mu=\bar c\gamma_\mu(1-\gamma_5)b,
\end{equation}
between the initial pseudoscalar $B_c$ meson and the relevant final-state meson. Defining
\begin{equation}
	q_\mu=(p-p^\prime)_\mu,\qquad P_\mu=(p+p^\prime)_\mu,
\end{equation}
the nonperturbative dynamics are encoded in Lorentz-invariant form factors that depend only on the momentum transfer $q^2$.

For transitions to pseudoscalar final states, such as $B_c\to\eta_c$, the vector-current matrix element is decomposed as
\begin{equation}
	\langle P(p^\prime)|\bar c\gamma_\mu b|B_c(p)\rangle
	=
	\left(P_\mu-\frac{m_{B_c}^2-m_P^2}{q^2}q_\mu\right)f_+(q^2)
	+
	\frac{m_{B_c}^2-m_P^2}{q^2}q_\mu\,f_0(q^2),
	\label{eq:ff_pp_main}
\end{equation}
with the kinematic constraint
\begin{equation}
	f_+(0)=f_0(0).
\end{equation}
The form factor $f_+(q^2)$ governs the dominant contribution in the light-lepton channels, while $f_0(q^2)$ becomes particularly important when the final lepton is a $\tau$.

For vector final states, such as $B_c\to J/\psi$, the hadronic matrix elements are expressed through four form factors:
\begin{equation}
	\langle V(p^\prime,\epsilon^\ast)|\bar c\gamma_\mu b|B_c(p)\rangle
	=
	\frac{2iV(q^2)}{m_{B_c}+m_V}\,
	\epsilon_{\mu\nu\alpha\beta}\,
	\epsilon^{\ast\nu}p^\alpha p^{\prime\,\beta},
	\label{eq:ff_pv_v_main}
\end{equation}
\begin{equation}
	\begin{aligned}
		\langle V(p^\prime,\epsilon^\ast)|\bar c\gamma_\mu\gamma_5 b|B_c(p)\rangle
		={}&
		2m_V\frac{\epsilon^\ast\!\cdot q}{q^2}q_\mu A_0(q^2)
		+(m_{B_c}+m_V)\left(\epsilon^\ast_\mu-\frac{\epsilon^\ast\!\cdot q}{q^2}q_\mu\right)A_1(q^2)
		\\
		&-
		\frac{\epsilon^\ast\!\cdot q}{m_{B_c}+m_V}
		\left(P_\mu-\frac{m_{B_c}^2-m_V^2}{q^2}q_\mu\right)A_2(q^2),
	\end{aligned}
	\label{eq:ff_pv_a_main}
\end{equation}
with the regularity condition
\begin{equation}
	2m_VA_0(0)=(m_{B_c}+m_V)A_1(0)-(m_{B_c}-m_V)A_2(0).
\end{equation}
These form factors encode the complete nonperturbative content of the semileptonic $B_c\to J/\psi$ transition.

For scalar final states such as $B_c\to\chi_{c0}$, the axial-current matrix element is parameterized as
\begin{equation}
	\langle S(p^\prime)|\bar c\gamma_\mu\gamma_5 b|B_c(p)\rangle
	=
	-i\left[
	\left(P_\mu-\frac{m_{B_c}^2-m_S^2}{q^2}q_\mu\right)F_1(q^2)
	+
	\frac{m_{B_c}^2-m_S^2}{q^2}q_\mu\,F_0(q^2)
	\right],
	\label{eq:ff_ps_main}
\end{equation}
with
\begin{equation}
	F_1(0)=F_0(0).
\end{equation}

For axial-vector final states, such as $B_c\to\chi_{c1}$ and $B_c\to h_c$, we use the standard decomposition in terms of $A$, $V_0$, $V_1$, and $V_2$,
\begin{equation}
	\langle A(p^\prime,\epsilon^\ast)|\bar c\gamma_\mu\gamma_5 b|B_c(p)\rangle
	=
	-\frac{2A(q^2)}{m_{B_c}-m_A}\,
	\epsilon_{\mu\nu\alpha\beta}\,
	\epsilon^{\ast\nu}p^\alpha p^{\prime\,\beta},
	\label{eq:ff_pa_a_main}
\end{equation}
\begin{equation}
	\begin{aligned}
		\langle A(p^\prime,\epsilon^\ast)|\bar c\gamma_\mu b|B_c(p)\rangle
		={}&
		-i\Bigg[
		(m_{B_c}-m_A)\epsilon^\ast_\mu V_1(q^2)
		-\frac{\epsilon^\ast\!\cdot p}{m_{B_c}-m_A}P_\mu V_2(q^2)
		\\
		&\qquad
		-2m_A\frac{\epsilon^\ast\!\cdot p}{q^2}q_\mu\big(V_3(q^2)-V_0(q^2)\big)
		\Bigg],
	\end{aligned}
	\label{eq:ff_pa_v_main}
\end{equation}
where
\begin{equation}
	V_3(q^2)=
	\frac{m_{B_c}-m_A}{2m_A}V_1(q^2)
	-
	\frac{m_{B_c}+m_A}{2m_A}V_2(q^2),
\end{equation}
and
\begin{equation}
	V_3(0)=V_0(0).
\end{equation}

For tensor final states, such as $B_c\to\chi_{c2}$, the matrix elements are written as
\begin{equation}
	\langle T(p^\prime,\epsilon^\ast)|\bar c\gamma_\mu b|B_c(p)\rangle
	=
	i\,h(q^2)\,
	\epsilon_{\mu\nu\rho\sigma}\,
	\epsilon^{\ast\nu\lambda}p_\lambda
	P^\rho q^\sigma,
	\label{eq:ff_pt_v_main}
\end{equation}
\begin{equation}
	\langle T(p^\prime,\epsilon^\ast)|\bar c\gamma_\mu\gamma_5 b|B_c(p)\rangle
	=
	k(q^2)\,\epsilon^\ast_{\mu\nu}p^\nu
	+\epsilon^\ast_{\alpha\beta}p^\alpha p^\beta
	\left[b_+(q^2)P_\mu+b_-(q^2)q_\mu\right].
	\label{eq:ff_pt_a_main}
\end{equation}
The same logic extends to the higher-spin $D$-wave charmonium channels, where the weak amplitudes are likewise reduced to finite sets of invariant form factors. In each case the nonperturbative problem is shifted to the determination of these scalar functions of $q^2$.

\subsection{Three-point QCD sum rules and analytic continuation}

The transition form factors introduced above are extracted from three-point correlation functions of the form
\begin{equation}
	\Pi_{\mu\cdots}(p^2,p^{\prime\,2},q^2)
	=
	i^2\!\int d^4x\,d^4y\,
	e^{ip^\prime\cdot x}e^{iq\cdot y}
	\langle0|T\{J_X(x)\,J_\mu(y)\,J_{B_c}^\dagger(0)\}|0\rangle,
	\label{eq:3ptcorr_main}
\end{equation}
where the Lorentz structure depends on the spin of the final-state meson. On the hadronic side, the correlator is expressed in terms of meson poles, decay constants, and transition form factors. On the QCD side, it is represented through a double dispersion relation,
\begin{equation}
	\Pi_i^{\rm QCD}(p^2,p^{\prime\,2},q^2)
	=
	\int ds\,du\,
	\frac{\rho_i^{\rm pert}(s,u,q^2)+\rho_i^{\rm nonpert}(s,u,q^2)}
	{(s-p^2)(u-p^{\prime\,2})}
	+\cdots,
	\label{eq:dispersion_main}
\end{equation}
where $\rho_i^{\rm pert}$ denotes the perturbative spectral density and $\rho_i^{\rm nonpert}$ contains the nonperturbative condensate contributions \cite{Cutkosky1960,ColangeloKhodjamirian2001}.

After matching the hadronic and QCD representations and applying the double Borel transformation, one obtains a sum rule for each invariant form factor. The continuum contribution is modeled through effective thresholds, while the Borel parameters are chosen so as to balance pole dominance against the convergence of the operator product expansion. The form factors are first obtained in the spacelike region, where the sum rules are reliable, and are then continued to the physical timelike region through a truncated $z$-series representation,
\begin{equation}
	F(q^2)=\frac{1}{1-q^2/m_{\rm pole}^2}
	\sum_{n=0}^{N-1} b_n
	\left[
	z(q^2,t_0)^n
	-(-1)^{n-N}\frac{n}{N}z(q^2,t_0)^N
	\right],
	\label{eq:zseries_main}
\end{equation}
with
\begin{equation}
	z(q^2,t_0)=
	\frac{\sqrt{t_+-q^2}-\sqrt{t_+-t_0}}
	{\sqrt{t_+-q^2}+\sqrt{t_+-t_0}},
	\qquad
	t_\pm=(m_{B_c}\pm m_X)^2.
\end{equation}
This parametrization preserves analyticity and provides a stable bridge between the spacelike sum-rule region and the physical semileptonic domain \cite{Bourrely2009BCL}.

\subsection{Helicity amplitudes and differential observables}

Once the invariant form factors are known, the semileptonic observables are constructed in the helicity basis. For pseudoscalar final states, the helicity amplitudes are
\begin{equation}
	H_0(q^2)=\frac{\sqrt{\lambda(m_{B_c}^2,m_P^2,q^2)}}{\sqrt{q^2}}\,f_+(q^2),
	\qquad
	H_t(q^2)=\frac{m_{B_c}^2-m_P^2}{\sqrt{q^2}}\,f_0(q^2),
	\label{eq:hel_pp_main}
\end{equation}
where
\begin{equation}
	\lambda(a,b,c)=a^2+b^2+c^2-2ab-2ac-2bc
\end{equation}
is the K\"all\'en function. The corresponding differential width is
\begin{equation}
	\frac{d\Gamma(B_c\to P\,\ell\bar\nu_\ell)}{dq^2}
	=
	\frac{G_F^2|V_{cb}|^2}{192\pi^3m_{B_c}^3}
	\lambda^{1/2}(m_{B_c}^2,m_P^2,q^2)
	\left(1-\frac{m_\ell^2}{q^2}\right)^2
	\left[
	\left(1+\frac{m_\ell^2}{2q^2}\right)|H_0|^2
	+\frac{3m_\ell^2}{2q^2}|H_t|^2
	\right].
	\label{eq:dg_pp_main}
\end{equation}

For vector and axial-vector final states, the helicity amplitudes are given by
\begin{equation}
	H_\pm(q^2)=
	(m_{B_c}+m_V)A_1(q^2)
	\mp
	\frac{\sqrt{\lambda(m_{B_c}^2,m_V^2,q^2)}}{m_{B_c}+m_V}V(q^2),
\end{equation}
\begin{equation}
	H_0(q^2)=
	\frac{
		(m_{B_c}^2-m_V^2-q^2)(m_{B_c}+m_V)A_1(q^2)
		-\lambda(m_{B_c}^2,m_V^2,q^2)A_2(q^2)/(m_{B_c}+m_V)
	}
	{2m_V\sqrt{q^2}},
\end{equation}
\begin{equation}
	H_t(q^2)=
	\frac{\sqrt{\lambda(m_{B_c}^2,m_V^2,q^2)}}{\sqrt{q^2}}A_0(q^2),
\end{equation}
and the differential width becomes
\begin{equation}
	\frac{d\Gamma(B_c\to V(A,T)\,\ell\bar\nu_\ell)}{dq^2}
	=
	\frac{G_F^2|V_{cb}|^2}{192\pi^3m_{B_c}^3}
	q^2\lambda^{1/2}
	\left(1-\frac{m_\ell^2}{q^2}\right)^2
	\left[
	\left(1+\frac{m_\ell^2}{2q^2}\right)
	\sum_{\lambda=\pm,0}|H_\lambda|^2
	+\frac{3m_\ell^2}{2q^2}|H_t|^2
	\right].
	\label{eq:dg_pv_main}
\end{equation}

From the same helicity amplitudes one constructs the forward-backward asymmetry,
\begin{equation}
	A_{FB}(q^2)=
	\frac{
		\displaystyle \int_0^1 d\cos\theta_\ell\,\frac{d^2\Gamma}{dq^2d\cos\theta_\ell}
		-
		\displaystyle \int_{-1}^0 d\cos\theta_\ell\,\frac{d^2\Gamma}{dq^2d\cos\theta_\ell}
	}
	{\displaystyle \frac{d\Gamma}{dq^2}},
\end{equation}
the longitudinal lepton polarization,
\begin{equation}
	P_L^\ell(q^2)=
	\frac{
		\displaystyle \frac{d\Gamma^{h_\ell=-1/2}}{dq^2}
		-
		\displaystyle \frac{d\Gamma^{h_\ell=+1/2}}{dq^2}
	}
	{\displaystyle \frac{d\Gamma}{dq^2}},
\end{equation}
and, for spinful final states, the longitudinal polarization fraction
\begin{equation}
	F_L(q^2)=\frac{d\Gamma_L/dq^2}{d\Gamma/dq^2}.
\end{equation}
The branching fraction and lepton-flavour-universality ratio are then written as
\begin{equation}
	\mathcal B(B_c\to X)=\tau_{B_c}\int dq^2\,\frac{d\Gamma}{dq^2},
\end{equation}
and
\begin{equation}
	R(X)=
	\frac{\Gamma(B_c\to X\,\tau\bar\nu_\tau)}
	{\Gamma(B_c\to X\,\mu\bar\nu_\mu)}.
\end{equation}
This form-factor-to-helicity construction is the key bridge between the QCD sum-rule calculation and the phenomenological observables discussed later.

\subsection{Purely leptonic, annihilation, and lifetime formulas}

For the charged pseudoscalar state, the purely leptonic width is given by
\begin{equation}
	\Gamma(P\to \ell\nu_\ell)=
	\frac{G_F^2}{8\pi}f_P^2|V_{cq/bq}|^2m_\ell^2
	\left(1-\frac{m_\ell^2}{m_P^2}\right)^2m_P,
	\label{eq:purelep_p_main}
\end{equation}
and for the vector partner by
\begin{equation}
	\Gamma(V\to \ell\bar\nu_\ell)=
	\frac{G_F^2}{12\pi}|V_{cq/bq}|^2f_V^2m_V^3
	\left(1-\frac{m_\ell^2}{m_V^2}\right)^2
	\left(1+\frac{m_\ell^2}{2m_V^2}\right).
	\label{eq:purelep_v_main}
\end{equation}
The branching fraction is then obtained from
\begin{equation}
	BR=\Gamma\times\tau.
	\label{eq:br_main}
\end{equation}

Within the spectator picture, the total width of the $B_c$ meson is written as
\begin{equation}
	\Gamma(B_c^+\to X)=\Gamma(b\to X)+\Gamma(c\to X)+\Gamma(\mathrm{Anni}),
	\label{eq:totalwidth_main}
\end{equation}
where the first two terms describe the $b$- and $c$-quark spectator decays and the third denotes weak annihilation of the $\bar bc$ pair \cite{ChangChen1994Decays,Gershtein1995PhysUsp,Kiselev2000,RaiVinodkumar2006}. Since these quantities depend directly on the decay constants and constituent masses, they provide an additional check on the internal consistency of the short-distance sector.

\subsection{Radiative-transition formalism}

Radiative transitions probe overlap integrals between different bound-state wave functions and therefore complement the weak sector. In the notation used below, $E1$ denotes an electric-dipole transition, which mainly changes the orbital angular momentum by one unit while conserving the total spin, whereas $M1$ denotes a magnetic-dipole transition, which mainly connects states with the same orbital angular momentum and different spin. The label ``1'' refers to the dipole order of the electromagnetic multipole. The electric-dipole transition width is calculated from
\begin{equation}
	\Gamma_{E1}\!\left(n^{2S+1}L_J\to n'{}^{2S+1}L'_{J'}\right)
	=
	\frac{4\alpha}{3}\langle e_Q\rangle^2
	\frac{E_\gamma^3E_f}{M_i}\,
	C_{fi}\,
	\delta_{SS'}
	\left|\left\langle n'{}^{2S+1}L'_{J'}|r|n^{2S+1}L_J\right\rangle\right|^2,
	\label{eq:e1_main}
\end{equation}
where
\begin{equation}
	\langle e_Q\rangle=\frac{m_Qe_q-m_qe_{\bar Q}}{m_Q+m_q},
	\qquad
	E_\gamma=\frac{M_i^2-M_f^2}{2M_i},
\end{equation}
and
\begin{equation}
	C_{fi}=\max(L,L')(2J'+1)
	\begin{Bmatrix}
		L' & J' & S\\
		J & L & 1
	\end{Bmatrix}.
\end{equation}

The magnetic-dipole transition width is written as
\begin{equation}
	\Gamma_{M1}(i\to f+\gamma)=
	\frac{16\alpha}{3}
	\left(\frac{m_{\bar q}e_Q-m_Qe_{\bar q}}{4m_{\bar q}m_Q}\right)^2
	E_\gamma^3(2J_f+1)
	\left|\left\langle f\left|j_0(E_\gamma r/2)\right|i\right\rangle\right|^2.
	\label{eq:m1_main}
\end{equation}
Because $E1$ and $M1$ transitions probe different combinations of orbital and spin structure, they provide a useful complementary test of the calculated wave functions.

\subsection{Regge-trajectory construction}

Finally, to summarize the global organization of the calculated spectrum, we construct Regge trajectories in the $(J,M^2)$ and $(n_r,M^2)$ planes using the linear relations
\begin{equation}
	J=\alpha M^2+\alpha_0,
	\label{eq:reggeJ_main}
\end{equation}
and
\begin{equation}
	n_r\equiv n-1=\beta M^2+\beta_0.
	\label{eq:reggenr_main}
\end{equation}
Here $M$ is the mass of the corresponding calculated or observed meson state, so $M^2$ is the squared meson mass, expressed in $\mathrm{GeV}^2$ in the numerical fits. The symbols $\alpha$ and $\beta$ are the fitted slopes and $\alpha_0$, $\beta_0$ are the corresponding intercepts. These trajectory fits do not replace the detailed spectroscopic calculation, but they provide a compact global characterization of whether the calculated $B_c$ states arrange themselves in the approximately linear fashion expected for heavy-meson spectroscopy \cite{Ebert2011Regge,DevlaniRai2014,Monteiro2017}. In this way, the Regge analysis serves as the broadest geometric summary of the same spectrum that enters the weak, radiative, and decay-constant sectors.

\section{QCD sum-rule inputs and numerical implementation}

\subsection{Three-point sum rules and Borel construction}

The weak-transition form factors are extracted from three-point correlation functions built from the interpolating currents of the initial and final mesons together with the charged weak current,
\begin{equation}
	\Pi_{\mu\cdots}(p,p^\prime)
	=
	i^2\int d^4x\,d^4y\,
	e^{ip^\prime\cdot x}e^{i(p-p^\prime)\cdot y}
	\langle0|T\{J_X(x)\,J_\mu(y)\,J_{B_c}^\dagger(0)\}|0\rangle,
	\label{eq:3ptcorr_compact}
\end{equation}
where the Lorentz structure depends on the quantum numbers of the final-state meson \(X\). For each channel, the correlator is decomposed into a complete basis of independent Lorentz tensors,
\begin{equation}
	\Pi_{\mu\cdots}(p,p^\prime)
	=
	\sum_i \Pi_i(p^2,p^{\prime\,2},q^2)\,L^{(i)}_{\mu\cdots}(p,p^\prime),
	\qquad q=p-p^\prime,
	\label{eq:lorentz_decomp_compact}
\end{equation}
so that each invariant amplitude \(\Pi_i\) can be matched separately between the hadronic and QCD representations. This is essential because the different Lorentz projections isolate different invariant form factors and need not display the same numerical stability.

On the hadronic side, the pole contribution of the lowest \(B_c\) and \(X\) states is written as
\begin{equation}
	\Pi_i^{\mathrm{had}}(p^2,p^{\prime\,2},q^2)
	=
	\frac{f_{B_c}m_{B_c}^{n_1}\,f_Xm_X^{n_2}}
	{(m_{B_c}^2-p^2)(m_X^2-p^{\prime\,2})}
	F_i(q^2)
	+\cdots,
	\label{eq:had_pole_compact}
\end{equation}
where \(F_i(q^2)\) denotes the corresponding transition form factor and the ellipsis represents excited states and continuum contributions. On the QCD side, the same amplitude is expressed through a double dispersion relation,
\begin{equation}
	\Pi_i^{\mathrm{QCD}}(p^2,p^{\prime\,2},q^2)
	=
	\int ds\int du\,
	\frac{\rho_i^{\mathrm{pert}}(s,u,q^2)+\rho_i^{\mathrm{nonpert}}(s,u,q^2)}
	{(s-p^2)(u-p^{\prime\,2})}
	+\Pi_i^{\mathrm{subtr}}(p^2,p^{\prime\,2},q^2),
	\label{eq:qcd_disp_compact}
\end{equation}
in which \(\rho_i^{\mathrm{pert}}\) arises from the perturbative triangle diagram and \(\rho_i^{\mathrm{nonpert}}\) contains the condensate corrections retained in the operator product expansion \cite{Cutkosky1960,ColangeloKhodjamirian2001}.

To suppress subtraction terms and enhance the ground-state contribution, we apply a double Borel transformation with respect to \(P^2=-p^2\) and \(P^{\prime\,2}=-p^{\prime\,2}\). This yields
\begin{equation}
	\widehat{\mathcal B}\,\Pi_i^{\mathrm{had}}
	=
	f_{B_c}m_{B_c}^{n_1}\,f_Xm_X^{n_2}\,
	F_i(q^2)\,
	e^{-m_{B_c}^2/T_1^2}e^{-m_X^2/T_2^2}
	+\cdots,
	\label{eq:borel_had_compact}
\end{equation}
and
\begin{equation}
	\widehat{\mathcal B}\,\Pi_i^{\mathrm{QCD}}
	=
	\int ds\int du\,
	\left[\rho_i^{\mathrm{pert}}(s,u,q^2)+\rho_i^{\mathrm{nonpert}}(s,u,q^2)\right]
	e^{-s/T_1^2}e^{-u/T_2^2}.
	\label{eq:borel_qcd_compact}
\end{equation}
Matching the two sides gives the working sum rule
\begin{equation}
	F_i(q^2)
	=
	\frac{e^{m_{B_c}^2/T_1^2}e^{m_X^2/T_2^2}}
	{f_{B_c}m_{B_c}^{n_1}\,f_Xm_X^{n_2}}
	\int_{s_{\mathrm{min}}}^{s_0}\!\!ds
	\int_{u_{\mathrm{min}}}^{u_0}\!\!du\,
	\left[\rho_i^{\mathrm{pert}}(s,u,q^2)+\rho_i^{\mathrm{nonpert}}(s,u,q^2)\right]
	e^{-s/T_1^2}e^{-u/T_2^2}.
	\label{eq:working_sumrule_compact}
\end{equation}

The effective continuum thresholds are parameterized as
\begin{equation}
	s_0=(m_{B_c}+\Delta_s)^2,
	\qquad
	u_0=(m_X+\Delta_u)^2,
	\label{eq:thresholds_compact}
\end{equation}
and the two Borel parameters are correlated through
\begin{equation}
	T_2^2=k\,T_1^2,
	\qquad
	k=\frac{m_X^2}{m_{B_c}^2},
	\label{eq:borel_corr_compact}
\end{equation}
or the equivalent relation used in the benchmark study of a given channel family. The working window is chosen by imposing pole dominance, acceptable convergence of the operator product expansion, and residual stability of the extracted form factor under variation of the Borel scale.

\subsection{Channel classes and sector-dependent normalization updates}

The numerical implementation is organized into four channel classes:
\begin{equation}
	B_c\to \eta_c,\ J/\psi;
	\qquad
	B_c\to \chi_{c0},\chi_{c1},h_c,\chi_{c2};
	\qquad
	B_c\to \psi_1,\psi_2,\eta_{c2},\psi_3;
	\qquad
	B_c\to D^{(*)}_{(s)}+\text{charmonium}.
\end{equation}
This division is not merely classificatory. The corresponding Lorentz decompositions, current structures, spectral densities, and numerical sensitivities differ substantially across these sectors, so the same formalism must be implemented with channel-specific thresholds, Borel windows, and normalization conventions \cite{Wu:2024gcq,Lu:2025bvi,Wu:2024dzn,Lu:2025usr}.

A central technical feature of the present analysis is that the weak-transition sector is updated to the heavy-quark masses employed in the spectroscopic part of the paper. Since the benchmark QCD sum-rule analyses do not all use the same quark-mass inputs, the resulting widths cannot be modified through one universal overall factor. Instead, the normalization is updated sector by sector. In the \(S\)-wave sector, the reference heavy-quark masses \(m_c^{(0)}\) and \(m_b^{(0)}\) enter the current normalization differently in the \(\eta_c\) and \(J/\psi\) channels, so the bare scaling factors are written separately as
\begin{equation}
	R_{\eta_c}^{\mathrm{bare}}
	=
	\frac{m_c(m_b+m_c)}{m_c^{(0)}(m_b^{(0)}+m_c^{(0)})},
	\qquad
	R_{J/\psi}^{\mathrm{bare}}
	=
	\frac{m_b+m_c}{m_b^{(0)}+m_c^{(0)}},
	\label{eq:swave_bare_ratios_compact}
\end{equation}
with the widths scaling as \((R^{\mathrm{bare}})^2\). After inclusion of the Coulomb-like enhancement inherited from the benchmark analysis, the corrected widths are written schematically as
\begin{equation}
	\Gamma_{\eta_c}^{\mathrm{corr}}
	=
	C_{\eta_c}\,\Gamma_{\eta_c}^{\mathrm{bare}},
	\qquad
	\Gamma_{J/\psi}^{\mathrm{corr}}
	=
	C_{J/\psi}\,\Gamma_{J/\psi}^{\mathrm{bare}},
	\label{eq:swave_corr_ratios_compact}
\end{equation}
so that the stronger normalization sensitivity of the pseudoscalar sector is preserved explicitly.

For the \(P\)- and \(D\)-wave sectors, the corresponding update is milder and is implemented through
\begin{equation}
	R_{\mathrm{P/D}}
	=
	\frac{m_b+m_c}{m_b^{(0)}+m_c^{(0)}},
	\label{eq:pd_ratio_compact}
\end{equation}
supplemented by any state-dependent factors required by the benchmark treatment of individual channels. This milder response is numerically important because it allows the \(P\)- and \(D\)-wave modes to serve as internal stability checks on the implementation. For the nonleptonic channels involving \(D^{(*)}_{(s)}\) mesons, the normalization update is assigned according to the dominant pseudoscalar or vector structure of the factorized amplitude, so that the distinction already seen in the pure semileptonic sector is maintained when open-charm mesons appear in the final state.

\subsection{Spacelike evaluation, analytic continuation, and helicity observables}

The sum rules are evaluated first in the spacelike domain,
\begin{equation}
	Q^2=-q^2>0,
\end{equation}
where the operator product expansion is most reliable and the three-point correlator is free from physical hadronic singularities. For each channel and for each invariant form factor, the sum rules are computed at a discrete set of spacelike points inside the accepted Borel window. These points are then fitted to the analyticity-based \(z\)-series form
\begin{equation}
	F(q^2)=\frac{1}{1-q^2/m_{\mathrm{pole}}^2}
	\sum_{n=0}^{N-1} b_n
	\left[
	z(q^2,t_0)^n
	-(-1)^{n-N}\frac{n}{N}z(q^2,t_0)^N
	\right],
	\label{eq:zfit_compact}
\end{equation}
with
\begin{equation}
	z(q^2,t_0)
	=
	\frac{\sqrt{t_+-q^2}-\sqrt{t_+-t_0}}
	{\sqrt{t_+-q^2}+\sqrt{t_+-t_0}},
	\qquad
	t_\pm=(m_{B_c}\pm m_X)^2.
	\label{eq:zvar_compact}
\end{equation}
The fitted coefficients \(b_n\) are constrained to satisfy the kinematic relations at \(q^2=0\), such as
\begin{equation}
	f_+(0)=f_0(0),\qquad F_1(0)=F_0(0),\qquad V_3(0)=V_0(0),
\end{equation}
together with the corresponding relation among \(A_0\), \(A_1\), and \(A_2\) in the vector channel.

Once the timelike form factors are reconstructed, they are inserted into the helicity-amplitude formulas of the previous section. In the pseudoscalar channels this means
\begin{equation}
	H_0(q^2)=\frac{\sqrt{\lambda}}{\sqrt{q^2}}\,f_+(q^2),
	\qquad
	H_t(q^2)=\frac{m_{B_c}^2-m_P^2}{\sqrt{q^2}}\,f_0(q^2),
\end{equation}
whereas the vector and axial-vector channels use the standard combinations of \(V\), \(A_0\), \(A_1\), and \(A_2\). The integrated semileptonic widths are then obtained by numerically integrating over
\begin{equation}
	m_\ell^2 \le q^2 \le (m_{B_c}-m_X)^2.
\end{equation}
From these widths we compute the branching fractions,
\begin{equation}
	\mathcal B(B_c\to X)=\tau_{B_c}\,\Gamma(B_c\to X),
\end{equation}
the lepton-flavour-universality ratios,
\begin{equation}
	R(X)=
	\frac{\Gamma(B_c\to X\,\tau\bar\nu_\tau)}
	{\Gamma(B_c\to X\,\mu\bar\nu_\mu)},
\end{equation}
and the integrated angular observables derived from the same helicity amplitudes, including the forward-backward asymmetry, longitudinal lepton polarization, and final-state longitudinal polarization fraction. Because all of these quantities are generated from a common fitted form-factor set, they provide a stringent internal check on the consistency of the analytic continuation.

\subsection{Implementation of the semileptonic and nonleptonic channels}

In practice, the \(S\)-wave channels
\begin{equation}
	B_c\to\eta_c,\qquad B_c\to J/\psi
\end{equation}
are evaluated first, since they provide the cleanest benchmark for the entire procedure. The invariant form factors \(f_+\), \(f_0\), \(V\), \(A_0\), \(A_1\), and \(A_2\) are extracted at spacelike \(Q^2\), fitted, and then used both for semileptonic decays and, after factorization, for the nonleptonic modes
\begin{equation}
	B_c^-\to \eta_c\,\pi^-,
	\ \eta_c\,K^-,
	\ \eta_c\,\rho^-,
	\ \eta_c\,K^{*-},
\end{equation}
\begin{equation}
	B_c^-\to J/\psi\,\pi^-,
	\ J/\psi\,K^-,
	\ J/\psi\,\rho^-,
	\ J/\psi\,K^{*-}.
\end{equation}
The same fitted form factors are therefore used consistently for total widths, branching fractions, and differential observables. The nonleptonic amplitudes are evaluated in the standard color-favoured factorization approximation,
\begin{equation}
\mathcal A(B_c^-\to X M^-)=\frac{G_F}{\sqrt{2}}\,V_{cb}V_{q_1q_2}^{\ast}\,a_1\,\langle M^-|\bar q_1\gamma^\mu(1-\gamma_5)q_2|0\rangle\,\langle X|\bar c\gamma_\mu(1-\gamma_5)b|B_c^-\rangle,
\label{eq:factorized_amp_main}
\end{equation}
where $X$ denotes the charmonium final state and $M^-$ denotes the emitted pseudoscalar or vector meson. For a pseudoscalar emitted meson one uses $\langle P^-(k)|A^\mu|0\rangle=i f_P k^\mu$, while for a vector emitted meson $\langle V^-(k,\varepsilon)|V^\mu|0\rangle=f_V m_V\varepsilon^{\ast\mu}$. The effective coefficient is $a_1=C_1+C_2/N_c$ in naive factorization, and the CKM factor is chosen according to the emitted meson, for example $V_{ud}^{\ast}$ for $\pi^-,\rho^-$ and $V_{us}^{\ast}$ for $K^-,K^{\ast-}$. The same formula is used for the open-charm modes by replacing the emitted light meson with $D_{(s)}^{(*)-}$ and using the corresponding decay constant and CKM factor.

The \(P\)-wave sector is implemented analogously, but with the larger invariant basis required by the scalar, axial-vector, and tensor channels. In particular, the scalar mode \(B_c\to\chi_{c0}\) uses \(F_0\) and \(F_1\), the axial-vector channels \(B_c\to\chi_{c1},h_c\) use \(A\), \(V_0\), \(V_1\), and \(V_2\), and the tensor channel uses the invariant functions associated with the projected tensor structures. The same formal procedure is applied to the \(D\)-wave channels, but in that sector the discussion focuses primarily on integrated observables because the available benchmark information is most robust at that level.

Finally, the factorized nonleptonic modes with open charm,
\begin{equation}
	B_c^-\to D^- \eta_c,\ D^- J/\psi,\ D^{*-}\eta_c,\ D^{*-}J/\psi,
\end{equation}
\begin{equation}
	B_c^-\to D_s^- \eta_c,\ D_s^- J/\psi,\ D_s^{*-}\eta_c,\ D_s^{*-}J/\psi,
\end{equation}
are evaluated through the two-body width formula
\begin{equation}
	\Gamma
	=
	\frac{|\bm p|}{8\pi m_{B_c}^2}\,|\mathcal A|^2,
\end{equation}
with \(\mathcal A\) given by Eq.~\eqref{eq:factorized_amp_main}, after summing over the vector-meson polarizations where required. Since these amplitudes mix transition form factors and meson decay constants, they provide a nontrivial check that the sector-dependent normalization update has been implemented consistently across both semileptonic and nonleptonic observables.

In this way, the numerical outputs presented later in the Results and Discussion section are generated from a single common chain: Lorentz projection of the three-point correlator, Borel-improved sum-rule extraction in the spacelike region, analytic continuation through the \(z\)-fit, sector-dependent normalization update, and helicity-based evaluation of the physical observables. This is the central technical logic underlying the weak sector of the present work.

\section{Results and discussion}

The results of the present study are most naturally interpreted as different consequences of one common screened-potential description of the \(B_c\) system. The mass spectrum fixes the radial and orbital scales, the corresponding wave functions determine the short-distance normalization, and the same inputs then enter the decay constants, weak widths, annihilation observables, radiative transitions, and Regge trajectories. For that reason, the discussion is organized by observable, with the corresponding tables and figures placed close to the relevant analysis. This arrangement makes the internal logic of the calculation explicit and allows direct comparison with the main theoretical and experimental benchmarks in each sector.

\subsection{Mass spectrum and spin splittings}

The mass spectrum provides the first test of the screened-potential framework. As seen in Tables~\ref{Table:MSA2}--\ref{Table:masses2b}, the low-lying \(S\)-wave states remain in the standard \(B_c\) region established by early and later quark-model analyses \cite{EichtenQuigg1994,Gershtein1995PRD,GuptaJohnson1996,Godfrey2004,Ebert2003Prop,RaiVinodkumar2006}. This is an important result, because it shows that the introduction of screening does not distort the well-established short-distance structure of the system.

\begin{table}[H]
	{\caption{\label{Table:MSA2}{Spin-averaged masses of the $B_c$ meson for $S$-, $P$-, $D$-, and $F$-wave states (in GeV). The column \(\xi\) gives the screening parameter used for each state.}}}
	\resizebox{\textwidth}{!}{
		\centering
		\begin{tabular}{c c c c c c c c c c c}
			\hline\noalign{\smallskip}
			\hline\noalign{\smallskip}
			Meson & State & $\xi$ & M$_{SA}$(This work) & Ref.\cite{DevlaniRai2014} & Ref.\cite{Ebert2003Prop} & Ref.\cite{RaiVinodkumar2006} & Ref.\cite{Godfrey2004} & Ref.\cite{Gershtein1995PRD} & Ref.\cite{Ebert2011Regge} & Ref.\cite{EichtenQuigg1994}\\
			\noalign{\smallskip}\hline\noalign{\smallskip}
			$B_{c}$ & 1S & 0.878 & 6.318 & 6.318 & 6.318 & 6.320 & 6.321 & 6.301 & 6.316 & 6.319 \\
			& 2S & 0.562 & 6.866 & 6.870 & 6.872 & 6.761 & 6.879 & 6.893 & 6.869 & 6.888\\
			& 3S & 0.477 & 7.226 & 7.248 & 7.250 & 7.085 & 7.266 & & 7.224 & 7.271 \\
			& 4S & 0.432 & 7.517 & 7.567 & 7.603 & & & & & \\
			& 5S & 0.401 & 7.767 & 7.854 & 7.942 & & & & & \\
			& 6S & 0.377 & 7.990 & 8.121 & 8.278 & & & & & \\
			
			\noalign{\smallskip}\hline\noalign{\smallskip}
			& 1P & 0.589 & 6.774 & 6.769 & 6.749 & 6.662 & 6.751 & 6.728 & 6.746 & 6.736 \\
			& 2P & 0.488 & 7.150 & 7.157 & 7.143 & 7.027 & 7.152 & 7.122 & 7.140 & 7.142 \\
			& 3P & 0.439 & 7.447 & 7.481 & 7.510 & & & & & \\
			& 4P & 0.406 & 7.702 & & & & & & & \\
			& 5P & 0.382 & 7.929 & & & & & & & \\
			
			\noalign{\smallskip}\hline\noalign{\smallskip}
			& 1D & 0.521 & 7.033 & 7.028 & 7.026 & & 7.039 & 7.008 & 7.078 & 7.009 \\
			& 2D & 0.458 & 7.365 & 7.365 & 7.400 & & & & & \\
			& 3D & 0.420 & 7.613 & 7.661 & 7.743 & & & & & \\
			& 4D & 0.392 & 7.847 & & & & & & & \\
			& 5D & 0.370 & 8.058 & & & & & & & \\
			
			\noalign{\smallskip}\hline\noalign{\smallskip}
			& 1F & 0.485 & 7.289 & & & & & & & \\
			& 2F & 0.438 & 7.518 & & & & & & & \\
			& 3F & 0.406 & 7.761 & & & & & & & \\
			& 4F & 0.381 & 7.979 & & & & & & & \\
			& 5F & 0.361 & 8.178 & & & & & & & \\
			\noalign{\smallskip}\hline\noalign{\smallskip}
		\end{tabular}
	}
\end{table}

The spin-averaged \(1S\) scale is also compatible with the lattice-QCD benchmarks of Refs.~\cite{Davies1996,Allison2005,Gregory2010,Dowdall2012}. Such agreement is particularly useful because the lattice calculations provide a non-model reference for the lowest \(B_c\) mass scale. The present framework therefore reproduces the accepted ground-state region not only within the spread of potential models, but also relative to the lattice expectation.

Beyond the ground state, the spectrum shows the characteristic effect of screening. The radial and orbital excitations remain within the broad quark-model band, but the higher \(S\)-, \(P\)-, \(D\)-, and \(F\)-wave levels are moderately compressed relative to many unscreened potentials. This trend is physically consistent with softened confinement and is closer in spirit to later studies that emphasize screened or modified upper-spectrum dynamics \cite{Ebert2011Regge,Li2019NRQM,Ortega2020,Li2023Survey,DevlaniRai2014,Abreu2020}. The present spectrum should therefore be viewed not simply as another interpolation among older potential-model results, but more specifically as a member of the class of \(B_c\) descriptions in which the low-lying states remain conventional while the upper spectrum is softened.

After inclusion of the spin-dependent terms, the resolved spectrum preserves the expected ordering and develops the standard fine and hyperfine structure. The pseudoscalar ground state remains close to the measured \(B_c^+\) mass, and the first radial excitation lies in the same broad region as the excited \(B_c\) signals reported by ATLAS, CMS, and LHCb \cite{LHCb2020Mass,ATLAS2014Excited,CMS2019Excited,LHCb2019Excited,PDG2024BflavoredHadrons}. In this respect, the model accommodates the available experimental anchors while retaining a coherent pattern for the unresolved higher states.

\begin{table}[H]
	\begin{center}
		{\caption{\label{Table:masses2a}{Resolved mass spectrum of the $B_c$ meson for $S$- and $P$-wave states (in GeV).}}}
		\resizebox{\textwidth}{!}{
			\begin{tabular}{ccccccccccccccc}
				\noalign{\smallskip}\hline\noalign{\smallskip}
				\noalign{\smallskip}\hline\noalign{\smallskip}
				$n^{2S+1}L_{J}$ & $J^P$ & This work & PDG \cite{PDG2024BflavoredHadrons}  & Ref.\cite{DevlaniRai2014}  & Ref.\cite{BhavinPatelVinodkumar2009}  &Ref.\cite{Ebert2003Prop}  &Ref.\cite{Ebert2011Regge}&Ref.\cite{Godfrey2004} &  Ref.\cite{Gershtein1995PRD} &Ref.\cite{EichtenQuigg1994}&Ref.\cite{ParmarPatelVinodkumar2010}& Ref.\cite{RaiVinodkumar2006}& Ref.\cite{AsgharAkramMasudSultan2019} & Ref.\cite{Abreu2020}\\
				\noalign{\smallskip}\hline\noalign{\smallskip}
				$1^1S_{0}$ & $0^{-}$  & 6.275 & $ 6.274 \pm 0.8  $ $(B^{+}_{c})$ & 6.278 & 6.275& 6.272 & 6.270 & 6.271& 6.253 & 6.264&6.256&6.269&6.318&6.277\\  
				$1^3S_{1}$  & $1^{-}$ & 6.332 &  & 6.331 & 6.331& 6.333 & 6.332 & 6.338& 6.317 & 6.337&6.343&6.337&6.336&6.288\\
				\noalign{\smallskip}
				$2^1S_{0}$ & $0^{-}$ & 6.858 &6.842 $(B_{c}(2S^{\pm})$\cite{ATLAS2014Excited,CMS2019Excited,LHCb2019Excited}& 6.853 & 6.852& 6.842 & 6.835 & 6.855& 6.867 & 6.856&6.939&6.743&6.741&6.845\\
				&&&6.871\cite{ATLAS2014Excited,CMS2019Excited,LHCb2019Excited}&\\
				&&&6.872\cite{ATLAS2014Excited,CMS2019Excited,LHCb2019Excited}&\\
				$2^3S_{1}$  & $1^{-}$ &6.868 &  & 6.873 & 6.997& 6.881 & 6.887 & 6.902& 6.899 & 6.890&&6.767&6.747&6.853\\
				\noalign{\smallskip}
				$3^1S_{0}$ & $0^{-}$ &7.223 &  &7.244 &7.236 &7.226 &7.193  &7.250 &&&&7.075 &7.014&7.284 \\ 
				$3^3S_{1}$ & $1^{-}$ &7.227 &   &7.249 &7.268 & 7.258 & 7.235 &7.272&&&&7.089 &7.018&7.290\\ 
				\noalign{\smallskip}
				$4^1S_{0}$ &$0^{-}$&7.514 &  &7.564 &7.542 &7.585  &  &&&&&&7.239  \\
				$4^3S_{1}$ &$1^{-}$& 7.517 &   &7.568 &7.570 &7.609  &  &&&&&&7.242 \\
				\noalign{\smallskip}
				$5^1S_{0}$ &$0^{-}$& 7.765 &  &7.852 &7.804 & 7.928 \\
				$5^3S_{1}$ &$1^{-}$& 7.767 &  &7.855 & 7.829&7.947  \\
				\noalign{\smallskip}
				$6^1S_{0}$ &$0^{-}$& 7.988 &  &8.120 &8.036 &  \\
				$6^3S_{1}$ &$1^{-}$& 7.990 &  &8.122 &8.060 &  \\
				\noalign{\smallskip}\hline\noalign{\smallskip}
				$1^3P_{0}$ &$0^{+}$&6.742&  & 6.748 & 6.770& 6.699 & 6.699 & 6.699& 6.683 & 6.700&6.848&6.642&6.631&6.639\\
				$1^{3}P_{1}$ &$1^{+}$&6.758&  & 6.767 & 6.781& 6.743 & 6.743 & 6.741& 6.717 & 6.730&6.882&6.662&6.650&6.606\\
				$1^{1}P_{1}$ & $1^{+}$ & 6.774&  & 6.769 & 6.793& 6.750 & 6.749 & 6.750& 6.729 & 6.736&6.882&6.672&6.656&6.656\\
				$1^3P_{2}$ & $2^{+}$ & 6.782&  & 6.775 & 6.804& 6.761 & 6.762 & 6.768& 6.743 & 6.747&6.889&6.650&6.665&6.667\\
				\noalign{\smallskip}
				$2^3P_{0}$ & $0^{+}$ &  7.126&  & 7.139 & 7.166& 7.094 & 7.091 & 7.122& 7.088 & 7.108&7.309&6.997&6.915&7.123\\
				$2^{3}P_{1}$ & $1^{+}$&  7.138&  & 7.155 & 7.176& 7.134 & 7.126 & 7.145& 7.113 & 7.135&7.337&7.012&6.930&7.088\\
				$2^{1}P_{1}$ &$1^{+}$ & 7.150&  & 7.156 & 7.185& 7.147 & 7.145 & 7.150& 7.124 & 7.142&7.339&7.027&6.939&7.121\\
				$2^3P_{2}$ & $2^{+}$ &  7.156&  & 7.162 & 7.194& 7.157 & 7.156 & 7.164& 7.134 & 7.153&7.345&7.042&6.946&7.127\\
				\noalign{\smallskip}
				$3^3P_{0}$ & $0^{+}$ &7.427 &  &7.463&&&&&&&&&7.147&7.523\\
				$3^{3}P_{1}$ & $1^{+}$ &7.437 &  &7.479&&&&&&&&&7.162&7.488\\
				$3^{1}P_{1}$ & $1^{+}$&7.447  & &7.479&&&&&&&&&7.168&7.513\\
				$3^3P_{2}$ & $2^{+}$ &7.452 & &7.785&&&&&&&&&7.176&7.515\\
				
				$4^3P_{0}$ & $0^{+}$ &7.684  &&&&&&&&&&&7.350\\
				$4^{3}P_{1}$ & $1^{+}$ &7.693   &&&&&&&&&&&7.364\\
				$4^{1}P_{1}$& $1^{+}$ & 7.702  &&&&&&&&&&&7.373\\
				$4^3P_{2}$ & $2^{+}$ & 7.707 &&&&&&&&&&&7.379\\
				
				$5^3P_{0}$ & $0^{+}$ & 7.875 &&&&&&\\
				$5^{3}P_{1}$ & $1^{+}$& 7.885 &&&&&&\\
				$5^{1}P_{1}$ & $1^{+}$& 7.896 &&&&&&\\
				$5^3P_{2}$ & $2^{+}$&7.901 &&&&&&\\
				\noalign{\smallskip}\hline\noalign{\smallskip}
			\end{tabular}
		}
	\end{center}
\end{table}

\begin{table}[H]
	\ContinuedFloat
	\begin{center}
		{\caption{\label{Table:masses2b}{Resolved mass spectrum of the $B_c$ meson for $D$- and $F$-wave states (continued, in GeV).}}}
		\resizebox{\textwidth}{!}{
			\begin{tabular}{ccccccccccccc}
				\noalign{\smallskip}\hline\noalign{\smallskip}
				\noalign{\smallskip}\hline\noalign{\smallskip}
				$n^{2S+1}L_{J}$ & $J^P$ & This work &  Ref.\cite{DevlaniRai2014}  & Ref.\cite{BhavinPatelVinodkumar2009}  &Ref.\cite{Ebert2003Prop}  &Ref.\cite{Ebert2011Regge}&Ref.\cite{Godfrey2004} &  Ref.\cite{Gershtein1995PRD} &Ref.\cite{EichtenQuigg1994}&Ref.\cite{ParmarPatelVinodkumar2010}&  Ref.\cite{AsgharAkramMasudSultan2019} & Ref.\cite{Abreu2020}\\
				\noalign{\smallskip}\hline\noalign{\smallskip}
				$1^3D_{1}$ &$1^{-}$& 7.030&   7.030 & 7.142& 7.021 & 7.021 & 7.028& 7.008 & 7.012&7.308&6.841&\\
				$1^{3}D_{2}$ &$2^{-}$& 7.033&   7.025 & 7.148& 7.025 & 7.025 & 7.036& 7.001 & 7.009&7.297&6.845&6.931\\
				$1^{1}D_{2}$ & $2^{-}$ &7.033&   7.035 & 7.155& 7.026 & 7.079 & 7.041& 7.016 & 7.012&7.304&6.845&6.920\\
				$1^3D_{3}$ & $3^{-}$ & 7.036&   7.026 & 7.146& 7.029 & 7.081 & 7.045& 7.007 & 7.005&7.286&6.847\\
				\noalign{\smallskip}
				$2^3D_{1}$ & $1^{-}$ & 7.361&   7.365 & & 7.392 &  & &  & &&7.080\\
				$2^{3}D_{2}$ & $2^{-}$& 7.366&   7.361 & & 7.399 &  & &  & &&7.084&7.334\\
				$2^{1}D_{2}$ &$2^{-}$ & 7.365&   7.370 & & 7.400 &  & &  & &&7.084&7.345\\
				$2^3D_{3}$ & $3^{-}$ &7.365&   7.363 & & 7.405 &  & &  & &&7.087\\
				\noalign{\smallskip}
				$3^3D_{1}$ & $1^{-}$ &7.615 &  7.664&&7.732&&&&&&7.289\\
				$3^{3}D_{2}$ & $2^{-}$ & 7.613&  7.656&&7.741&&&&&&7.293&7.694\\
				$3^{1}D_{2}$ & $2^{-}$&7.614  &  7.668&&7.743&&&&&&7.284&7.704\\
				$3^3D_{3}$ & $3^{-}$ & 7.615 &  7.659&&7.750&&&&&&7.287\\
				\noalign{\smallskip}
				$4^3D_{1}$ & $1^{-}$ &7.845  &&&&&&&&&7.478\\
				$4^{3}D_{2}$ & $2^{-}$ &7.847  &&&&&&&&&7.482\\
				$4^{1}D_{2}$ & $2^{-}$ &7.848 &&&&&&&&&7.483\\
				$4^3D_{3}$ & $3^{-}$ &7.849 &&&&&&&&&7.489\\
				\noalign{\smallskip}
				$5^3D_{1}$ & $1^{-}$ &8.057 &&&&&\\
				$5^{3}D_{2}$ & $2^{-}$&8.058 &&&&&\\
				$5^{1}D_{2}$ & $2^{-}$& 8.059 &&&&&\\
				$5^3D_{3}$ & $3^{-}$&8.060 &&&&&\\
				\noalign{\smallskip}\hline\noalign{\smallskip}
				$1^3F_{2}$ &$2^{+}$&7.236 &&&&&&&&&6.997\\
				$1^{3}F_{3}$ &$3^{+}$&7.239&&&&&&&&&6.994\\
				$1^{1}F_{3}$  & $3^{+}$ &7.239 &&&&&&&&&7.001\\
				$1^3F_{4}$ & $4^{+}$ & 7.245&&&&&&&&&6.996\\
				\noalign{\smallskip}
				$2^3F_{2}$ & $2^{+}$& 7.515&&&&&&&&&7.212\\
				$2^{3}F_{3}$& $3^{+}$& 7.518&&&&&&&&&7.211\\
				$2^{1}F_{3}$  &$3^{+}$&7.518&&&&&&&&&7.214 \\
				$2^3F_{4}$ & $4^{+}$ & 7.523&&&&&&&&&7.212\\
				\noalign{\smallskip}
				$3^3F_{2}$ & $2^{+}$ & 7.759 &  &&&&\\
				$3^{3}F_{3}$ & $3^{+}$ & 7.761  &&&&\\
				$3^{1}F_{3}$  & $3^{+}$& 7.761 &  &&&&\\
				$3^3F_{4}$ & $4^{+}$ &7.765 &  &&&&\\
				\noalign{\smallskip}
				$4^3F_{2}$ & $2^{+}$ &7.977 &&&&&\\
				$4^{3}F_{3}$ & $3^{+}$ &7.979 &&&&&\\
				$4^{1}F_{3}$  & $3^{+}$ &7.979 &&&&&\\
				$4^3F_{4}$ & $4^{+}$ &7.982 &&&&&\\
				\noalign{\smallskip}
				$5^3F_{2}$ & $2^{+}$ & 8.176 &&&&&\\
				$5^{3}F_{3}$ & $3^{+}$&8.178  &&&&&\\
				$5^{1}F_{3}$ & $3^{+}$&8.178  &&&&&\\
				$5^3F_{4}$ & $4^{+}$&8.181 &&&&&\\
				\noalign{\smallskip}\hline\noalign{\smallskip}
			\end{tabular}
		}
	\end{center}
\end{table}

A useful way to read Tables~\ref{Table:MSA2}--\ref{Table:masses2b} is to distinguish three comparison levels. For the lowest states, experiment, PDG averages, and lattice-QCD benchmarks are the primary reference points, and the present results are compatible with them. For the low- and intermediate-excitation region, the relevant comparison set is the established relativistic and constituent quark-model literature, where the present masses remain within the usual spread. For the higher radial and orbital excitations, the more meaningful comparison is with recent screened, softened, or coupled-channel studies, since these approaches address the same part of the spectrum most directly \cite{Ebert2011Regge,Li2019NRQM,Patnaik2024Review,Ortega2020,Li2023Survey,MartinGonzalez2022}. Under this three-level comparison, the present framework is conservative in the low-lying sector and more distinctive in the upper spectrum, where screening effects are expected to be most visible.

This point is important for the rest of the phenomenology. The weak, radiative, and Regge analyses all use the same spectroscopic input, so the quality of the mass spectrum is not an isolated issue. The broad agreement with both classic and modern spectroscopy studies therefore strengthens the interpretation of the subsequent decay and trajectory results as consequences of a realistic underlying bound-state framework.

\subsection{Decay constants}

The decay constants probe the short-distance behavior of the \(B_c\) wave functions and provide a direct bridge between the spectroscopic sector and the weak-decay analysis. Since both \(f_P\) and \(f_V\) are controlled mainly by the wave function at the origin, they offer a compact test of whether the model yields a reasonable short-distance normalization. In the present calculation, both constants decrease monotonically with radial excitation, as expected for progressively more extended \(nS\) states.

\begin{table}[H]
	\begin{center}
		{\caption{\label{Table:decay}{Decay constants of the pseudoscalar and vector $B_c$ mesons (in GeV). Representative comparisons are taken from Refs.~\cite{Ebert2003Prop,RaiVinodkumar2006,DevlaniRai2014,SunNiChen2023}; see also Refs.~\cite{Patnaik2024Review,Baker2014,Feng2023ThreeLoop,Kiselev2004Leptonic}.}}}
		\begin{tabular}{c c c c c c c}
			\noalign{\smallskip}\hline\noalign{\smallskip}
			\noalign{\smallskip}\hline\noalign{\smallskip}
			Meson & State & 1S & 2S & 3S & 4S & 5S \\
			\noalign{\smallskip}\hline\noalign{\smallskip}
			& $f_{P,\mathrm{cor}}$ & 0.410 & 0.165 & 0.112 & 0.088 & 0.073 \\
			& $f_{P}$ & 0.479 & 0.192 & 0.131 & 0.102 & 0.085 \\
			$B_{c}$ & Ref.~\cite{DevlaniRai2014} & 0.469 & 0.289 & 0.256 & 0.241 & 0.230 \\
			& Ref.~\cite{RaiVinodkumar2006} & 0.525 &  &  &  &  \\
			& Ref.~\cite{Ebert2003Prop} & 0.433 &  &  &  &  \\
			\noalign{\smallskip}\hline\noalign{\smallskip}
			& $f_{V,\mathrm{cor}}$ & 0.412 & 0.165 & 0.112 & 0.088 & 0.073 \\
			& $f_{V}$ & 0.481 & 0.192 & 0.131 & 0.102 & 0.085 \\
			$B_{c}^{*}$ & Ref.~\cite{DevlaniRai2014} & 0.471 & 0.290 & 0.257 & 0.241 & 0.231 \\
			& Ref.~\cite{RaiVinodkumar2006} & 0.528 &  &  &  &  \\
			& Ref.~\cite{Ebert2003Prop} & 0.503 &  &  &  &  \\
			\noalign{\smallskip}\hline\noalign{\smallskip}
		\end{tabular}
	\end{center}
\end{table}

The values listed in Table~\ref{Table:decay} place the ground-state decay constants within the standard theoretical range. In particular, the uncorrected value of \(f_P\) lies close to the results of Ebert \textit{et al.} and Devlani \textit{et al.}, while the corrected value moves toward the more conservative range favored by modern QCD-based estimates \cite{Ebert2003Prop,Baker2014,DevlaniRai2014,SunNiChen2023,Colquhoun2015}. This behavior is encouraging, because it indicates that the screened-potential framework reproduces a realistic short-distance normalization without driving \(f_{B_c}\) to the upper edge of the theoretical spread.

Another notable feature is the near equality \(f_V\simeq f_P\) in the low-lying sector. This reflects the heavy-heavy character of the \(B_c\) system and the comparatively modest breaking of spin symmetry in local couplings. The same pattern is consistent with relativistic quark-model expectations and with lattice-QCD indications that \(f_{B_c^\ast}/f_{B_c}\) is close to unity \cite{Ebert2003Prop,Colquhoun2015}. In the present context, this is also compatible with the small \(S\)-wave hyperfine splitting obtained in the mass spectrum.

The excited-state trend is more distinctive. Relative to some earlier quark-model calculations, the present results show a stronger suppression beyond the \(2S\) level, especially when compared with Ref.~\cite{DevlaniRai2014}. This is a natural consequence of screening: the same dynamics that compress the upper mass spectrum also reduce the short-distance overlap of higher radial states. The decay constants therefore support the central picture that the low-lying \(B_c\) states remain conventional, while higher excitations carry more visible signatures of screened-potential dynamics.

Overall, the decay-constant sector supports the model at both the ground-state and excited-state levels. The \(1S\) values remain compatible with the main phenomenological, lattice, and sum-rule benchmarks, while the stronger suppression of the higher \(nS\) states emerges as a specific prediction of the present framework.

\subsection{Weak decays to charmonia}

The weak-decay analysis reveals a clear hierarchy across the \(S\)-, \(P\)-, and \(D\)-wave charmonium sectors. The updated heavy-quark masses and the Coulomb-corrected treatment significantly increase the overall normalization of the widths relative to the bare calculation, but they do not alter the basic ordering of channels. This stability is one of the main results of the present study: the normalization is input-sensitive, whereas the spin-orbital hierarchy of the transitions is robust. At the qualitative level, this is consistent with the pattern found in lattice-QCD, QCD sum-rule, perturbative-QCD, Bethe--Salpeter, and relativistic quark-model studies, which generally place the dominant \(B_c\) weak transitions in the low-lying \(S\)-wave sector and find increasing suppression as the final-state orbital structure becomes more complicated \cite{Lu:2025bvi,Lu:2025usr,Lytle2016BcLattice,Harrison2025BcJpsiFF,Hu2019BcPQCDLat,Biswas2025BcS,Issadykov2018CCQM,Wang2011BcPWaveBS,Akan2022BcPWaveLCSR,Chen2021BcPwaveNLO,Ivanov2005BcCharmonia}.

\begin{figure}[H]
	\centering
	\begin{subfigure}{0.48\textwidth}
		\centering
		\includegraphics[width=\linewidth]{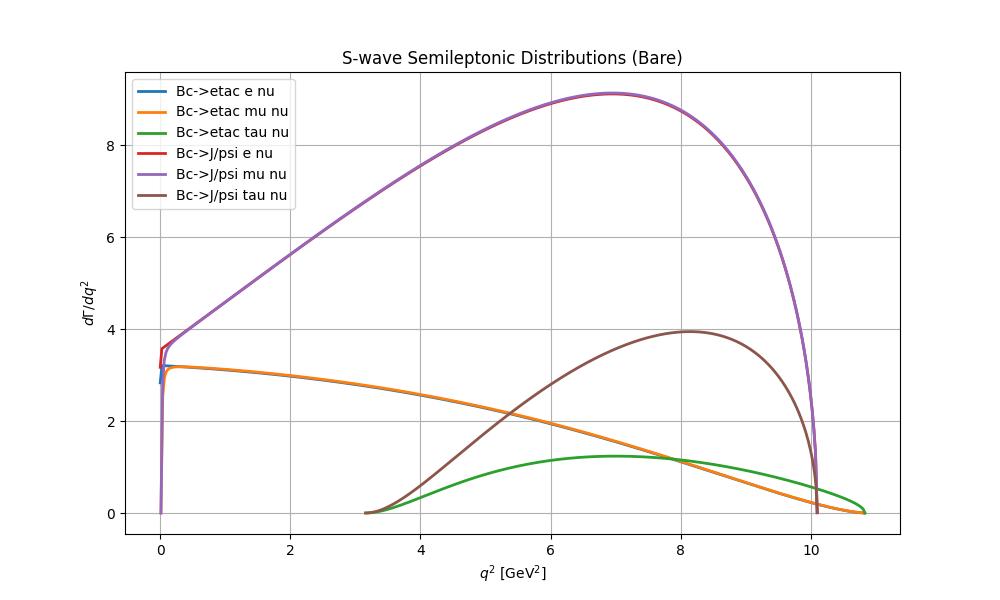}
	\end{subfigure}
	\hfill
	\begin{subfigure}{0.48\textwidth}
		\centering
		\includegraphics[width=\linewidth]{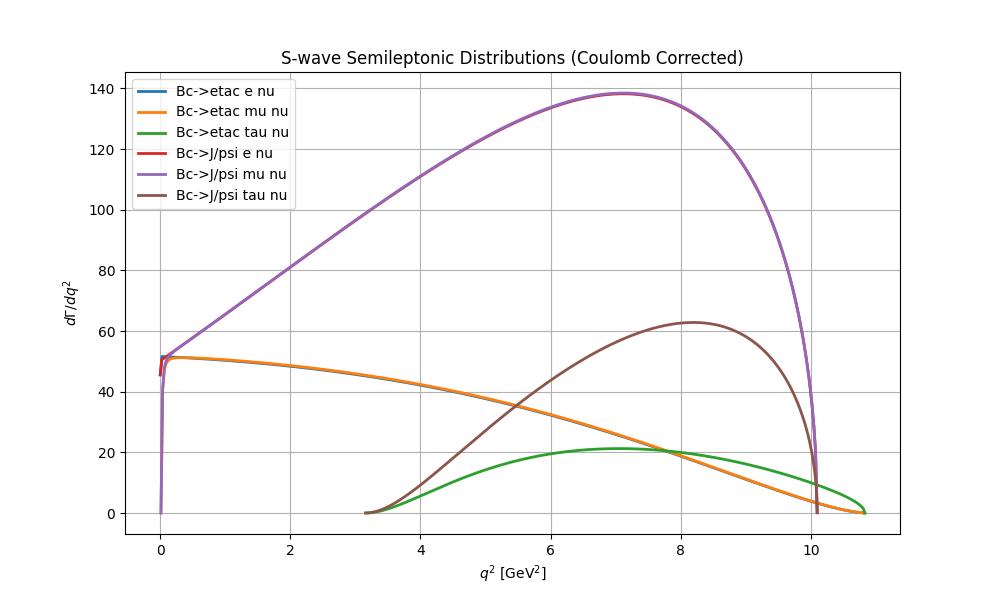}
	\end{subfigure}
	\caption{Decay-width distributions in the \(S\)-wave sector for the bare and Coulomb-corrected variants.}
	\label{fig:swave_dist}
\end{figure}

\begin{figure}[H]
	\centering
	\begin{subfigure}{0.48\textwidth}
		\centering
		\includegraphics[width=\linewidth]{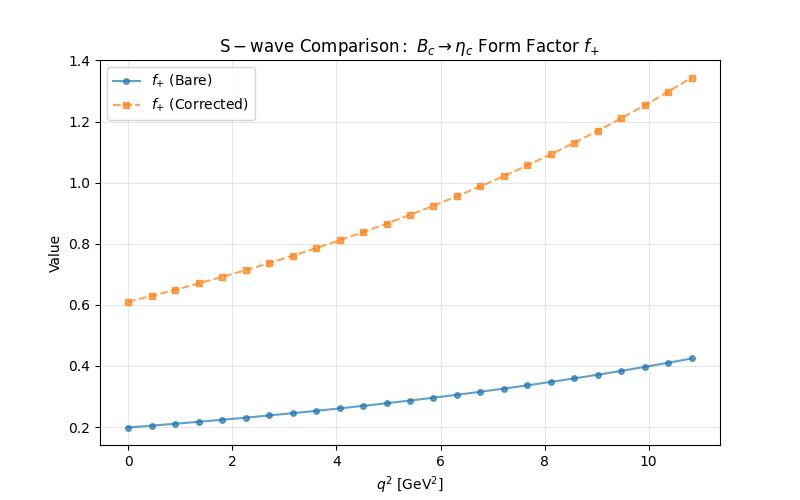}
	\end{subfigure}
	\hfill
	\begin{subfigure}{0.48\textwidth}
		\centering
		\includegraphics[width=\linewidth]{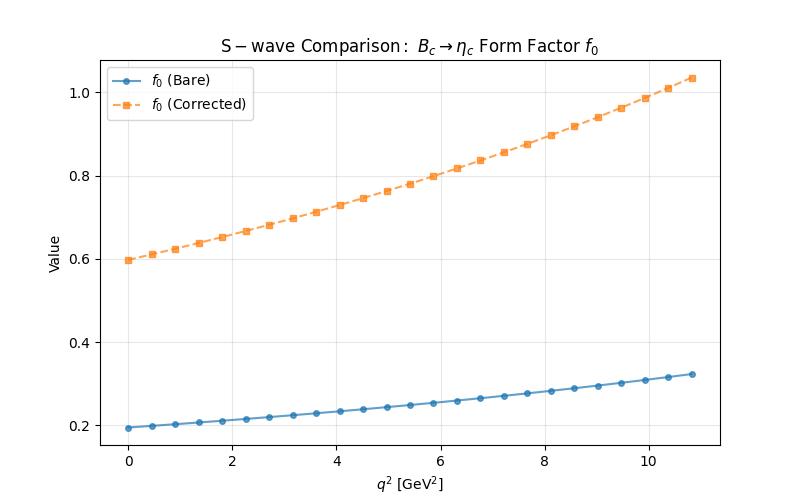}
	\end{subfigure}
	\caption{Bare and Coulomb-corrected form factors \(f_+(q^2)\) and \(f_0(q^2)\) for \(B_c\to\eta_c\).}
	\label{fig:swave_ff_etac}
\end{figure}

\begin{figure}[H]
	\centering
	\begin{subfigure}{0.48\textwidth}
		\centering
		\includegraphics[width=\linewidth]{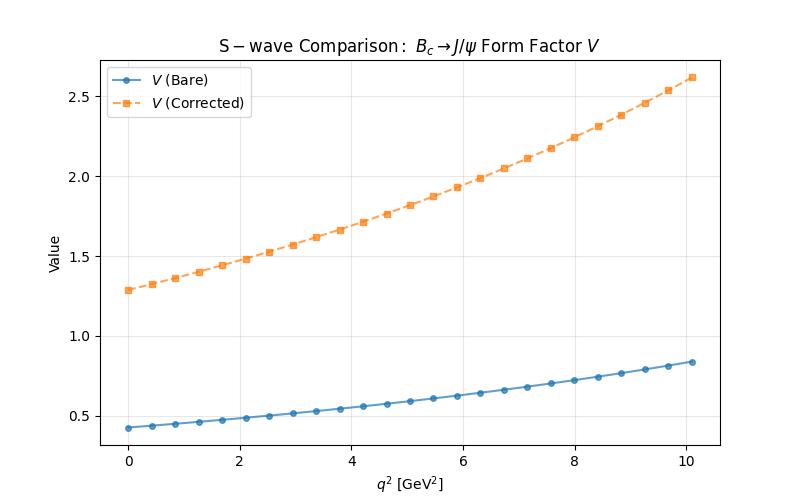}
	\end{subfigure}
	\hfill
	\begin{subfigure}{0.48\textwidth}
		\centering
		\includegraphics[width=\linewidth]{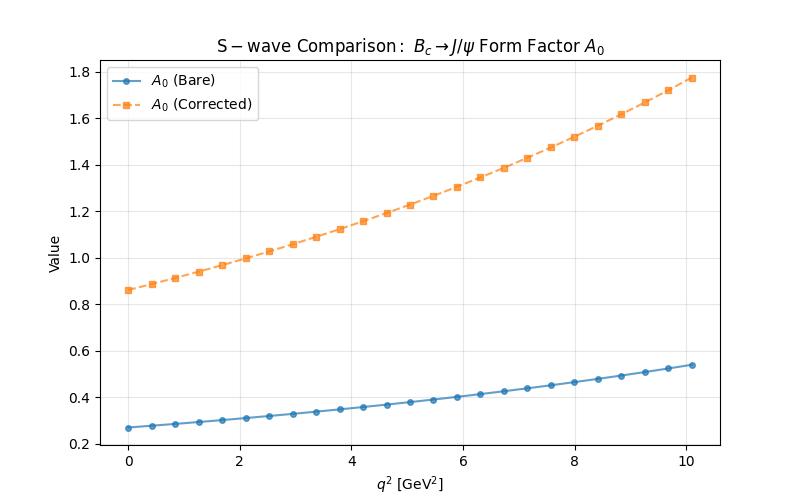}
	\end{subfigure}
	
	\vspace{0.8em}
	
	\begin{subfigure}{0.48\textwidth}
		\centering
		\includegraphics[width=\linewidth]{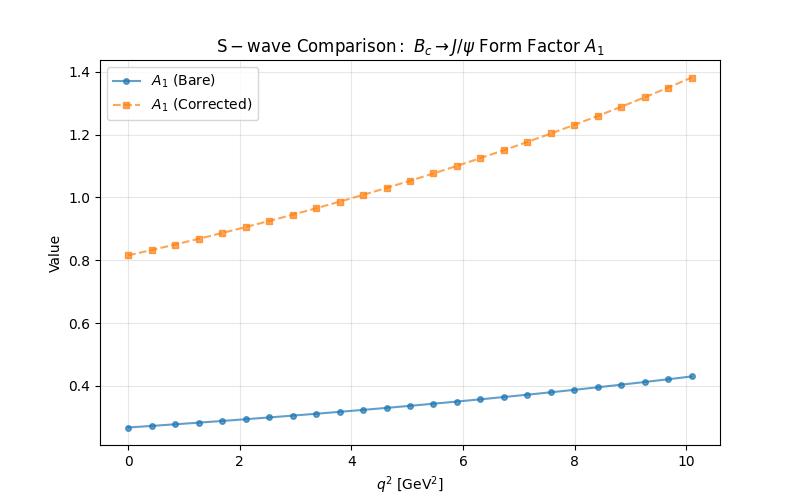}
	\end{subfigure}
	\hfill
	\begin{subfigure}{0.48\textwidth}
		\centering
		\includegraphics[width=\linewidth]{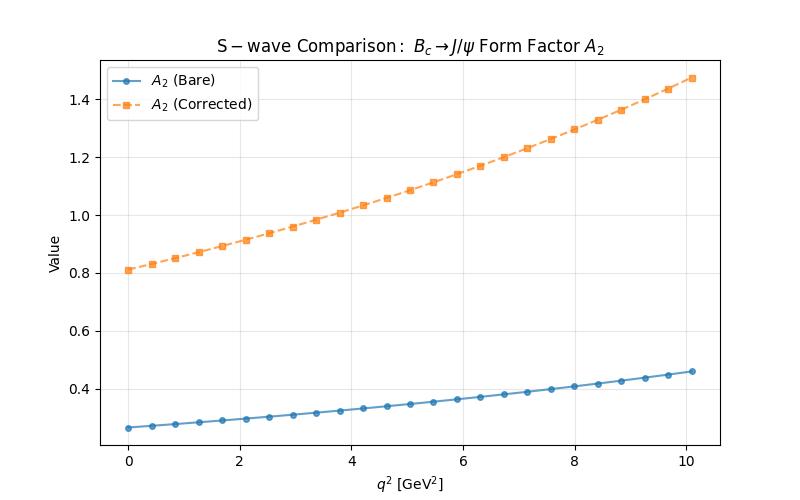}
	\end{subfigure}
	\caption{Bare and Coulomb-corrected form factors \(V(q^2)\), \(A_0(q^2)\), \(A_1(q^2)\), and \(A_2(q^2)\) for \(B_c\to J/\psi\).}
	\label{fig:swave_ff_jpsi}
\end{figure}

In the \(S\)-wave sector, Tables~\ref{tab:SWaveWidths}--\ref{tab:S_q2_psi} and Figs.~\ref{fig:swave_dist}--\ref{fig:swave_ff_jpsi} show that the channels involving \(\eta_c\) and \(J/\psi\) remain the dominant exclusive weak modes. The corrected \(\eta_c\) widths increase more strongly than the corresponding \(J/\psi\) widths, indicating that the update is not a uniform rescaling but acts through the detailed form-factor structure. Even so, the \(J/\psi\) channels continue to dominate the semileptonic sector in absolute size. This ordering is in line with full-lattice and hybrid lattice/HQSS analyses, which likewise identify \(B_c\to J/\psi\ell\bar\nu_\ell\) as one of the leading semileptonic channels and find a more moderate \(J/\psi\)-to-\(\eta_c\) splitting at the level of ratios than of raw normalizations \cite{Lytle2016BcLattice,Harrison2025BcJpsiFF,Hu2019BcPQCDLat,Biswas2025BcS,Issadykov2018CCQM}.

\begin{longtable}{>{\raggedright\arraybackslash}p{5.4cm}cccc}
	\caption{\label{tab:SWaveWidths}S-wave decay widths. This work is compared with Ref.~\cite{Wu:2024gcq}. All widths are in units of \(10^{-7}~\mathrm{eV}\).}\\
	\toprule
	Decay & This work (bare) & Ref.~\cite{Wu:2024gcq} (bare) & This work (corr.) & Ref.~\cite{Wu:2024gcq} (corr.)\\
	\midrule
	\endfirsthead
	\toprule
	Decay & This work (bare) & Ref.~\cite{Wu:2024gcq} (bare) & This work (corr.) & Ref.~\cite{Wu:2024gcq} (corr.)\\
	\midrule
	\endhead
	\midrule
	\multicolumn{5}{r}{Continued on next page}\\
	\endfoot
	\bottomrule
	\endlastfoot
	$\Bc^- \to \etac \pi^-$ & 3.403 & 1.8 & 55.334 & 17\\
	$\Bc^- \to \etac K^-$ & 0.284 & 0.15 & 4.557 & 1.4\\
	$\Bc^- \to \etac \rho^-$ & 9.264 & 4.9 & 149.727 & 46\\
	$\Bc^- \to \etac K^{*-}$ & 0.492 & 0.26 & 8.137 & 2.5\\
	$\Bc^- \to \etac e\bar\nu_e$ & 20.797 & 11 & 341.768 & 105\\
	$\Bc^- \to \etac \mu\bar\nu_\mu$ & 20.797 & 11 & 341.768 & 105\\
	$\Bc^- \to \etac \tau\bar\nu_\tau$ & 6.239 & 3.3 & 107.413 & 33\\
	$\Bc^- \to \psij \pi^-$ & 3.833 & 3.3 & 57.490 & 30\\
	$\Bc^- \to \psij K^-$ & 0.302 & 0.26 & 4.408 & 2.3\\
	$\Bc^- \to \psij \rho^-$ & 12.776 & 11 & 183.967 & 96\\
	$\Bc^- \to \psij K^{*-}$ & 0.697 & 0.60 & 10.348 & 5.4\\
	$\Bc^- \to \psij e\bar\nu_e$ & 70.848 & 61 & 1061.641 & 554\\
	$\Bc^- \to \psij \mu\bar\nu_\mu$ & 70.848 & 61 & 1057.808 & 552\\
	$\Bc^- \to \psij \tau\bar\nu_\tau$ & 17.422 & 15 & 275.950 & 144\\
\end{longtable}

\begin{table}[H]
	\centering
	\caption{Branching fractions for the \(S\)-wave sector in this work, in units of \(10^{-3}\).}
	\label{tab:SWaveBR}
	\begin{tabular}{lcc}
		\toprule
		Decay channel & This work (bare) & This work (corrected) \\
		\midrule
		$B_c^- \to \eta_c \pi^-$            & 0.264 & 4.287 \\
		$B_c^- \to \eta_c K^-$              & 0.022 & 0.353 \\
		$B_c^- \to \eta_c \rho^-$           & 0.718 & 11.601 \\
		$B_c^- \to \eta_c K^{*-}$           & 0.038 & 0.631 \\
		$B_c^- \to \eta_c e\bar\nu_e$       & 1.611 & 26.481 \\
		$B_c^- \to \eta_c \mu\bar\nu_\mu$   & 1.611 & 26.481 \\
		$B_c^- \to \eta_c \tau\bar\nu_\tau$ & 0.483 & 8.323 \\
		$B_c^- \to J/\psi \pi^-$            & 0.297 & 4.455 \\
		$B_c^- \to J/\psi K^-$              & 0.023 & 0.342 \\
		$B_c^- \to J/\psi \rho^-$           & 0.990 & 14.254 \\
		$B_c^- \to J/\psi K^{*-}$           & 0.054 & 0.802 \\
		$B_c^- \to J/\psi e\bar\nu_e$       & 5.490 & 82.259 \\
		$B_c^- \to J/\psi \mu\bar\nu_\mu$   & 5.490 & 81.962 \\
		$B_c^- \to J/\psi \tau\bar\nu_\tau$ & 1.350 & 21.381 \\
		\bottomrule
	\end{tabular}
\end{table}

\begin{table}[H]
	\centering
	\caption{Integrated angular and polarization observables in the \(S\)-wave sector.}
	\label{tab:SWaveIntegrated}
	\begin{tabular}{llcccc}
		\toprule
		Channel & Mode & $\langle A_{FB}\rangle_{\rm bare}$ & $\langle A_{FB}\rangle_{\rm corr}$ & $\langle P_L^\ell\rangle_{\rm bare}$ & $\langle P_L^\ell\rangle_{\rm corr}$ \\
		\midrule
		\multirow{2}{*}{$B_c\to\eta_c\ell\bar\nu_\ell$}
		& $\mu$  & -0.0128 & -0.0127 & -0.9643 & -0.9647 \\
		& $\tau$ & -0.3625 & -0.3613 & +0.3319 & +0.3385 \\
		\midrule
		\multirow{2}{*}{$B_c\to J/\psi\,\ell\bar\nu_\ell$}
		& $\mu$  & +0.2340 & +0.2303 & -0.9877 & -0.9875 \\
		& $\tau$ & +0.2968 & +0.2944 & -0.5099 & -0.5049 \\
		\bottomrule
	\end{tabular}
\end{table}

\begin{table}[H]
	\centering
	\caption{Integrated \(J/\psi\) longitudinal polarization fraction in the \(S\)-wave sector.}
	\label{tab:SWaveFL}
	\begin{tabular}{lcc}
		\toprule
		Mode & $\langle F_L\rangle_{\rm bare}$ & $\langle F_L\rangle_{\rm corr}$ \\
		\midrule
		$B_c\to J/\psi\,\mu\bar\nu_\mu$   & 0.4922 & 0.4898 \\
		$B_c\to J/\psi\,\tau\bar\nu_\tau$ & 0.4389 & 0.4413 \\
		\bottomrule
	\end{tabular}
\end{table}

\begin{table}[H]
	\centering
	\caption{Representative corrected \(q^2\)-dependent observables for \(B_c\to\eta_c\ell\bar\nu_\ell\). Differential widths are in units of \(10^{-7}\,\mathrm{eV/GeV^2}\).}
	\label{tab:S_q2_eta}
	\begin{tabular}{llcccc}
		\toprule
		Mode & $q^2$ (GeV$^2$) & $d\Gamma/dq^2$ & $A_{FB}(q^2)$ & $P_L^\ell(q^2)$ \\
		\midrule
		\multirow{3}{*}{$\mu$}
		& 2.70 & 46.79 & -0.00666 & -0.9814 \\
		& 5.41 & 35.77 & -0.00392 & -0.9879 \\
		& 8.12 & 17.99 & -0.00353 & -0.9864 \\
		\midrule
		\multirow{3}{*}{$\tau$}
		& 3.79 & 3.68  & -0.4683 & +0.3470 \\
		& 6.49 & 20.78 & -0.3820 & +0.2521 \\
		& 9.20 & 15.09 & -0.3098 & +0.4498 \\
		\bottomrule
	\end{tabular}
\end{table}

\begin{table}[H]
	\centering
	\caption{Representative corrected \(q^2\)-dependent observables for \(B_c\to J/\psi\,\ell\bar\nu_\ell\). Differential widths are in units of \(10^{-7}\,\mathrm{eV/GeV^2}\).}
	\label{tab:S_q2_psi}
	\begin{tabular}{llccccc}
		\toprule
		Mode & $q^2$ (GeV$^2$) & $d\Gamma/dq^2$ & $A_{FB}(q^2)$ & $P_L^\ell(q^2)$ & $F_L(q^2)$ \\
		\midrule
		\multirow{3}{*}{$\mu$}
		& 2.51 & 89.80  & +0.2225 & -0.9881 & 0.6241 \\
		& 5.03 & 125.16 & +0.2773 & -0.9957 & 0.4668 \\
		& 7.54 & 137.94 & +0.2496 & -0.9979 & 0.3832 \\
		\midrule
		\multirow{3}{*}{$\tau$}
		& 3.52 & 2.30  & +0.4482 & +0.0221 & 0.6787 \\
		& 6.03 & 44.88 & +0.3623 & -0.3841 & 0.4999 \\
		& 8.54 & 62.22 & +0.2471 & -0.6263 & 0.3836 \\
		\bottomrule
	\end{tabular}
\end{table}

The integrated observables strengthen this interpretation. From the corrected results one obtains \(R(\eta_c)\simeq0.314\) and \(R(J/\psi)\simeq0.261\), both consistent with recent Standard Model benchmarks. In particular, \(R(\eta_c)\) agrees well with the recent HQSS-based and pQCD+lattice estimates, while \(R(J/\psi)\) is essentially identical to the updated HPQCD lattice result and remains within the standard theoretical range \cite{Harrison2025BcJpsiFF,Hu2019BcPQCDLat,Biswas2025BcS}. The polarization observables are similarly satisfactory: \(\langle F_L\rangle(B_c\to J/\psi\tau\bar\nu_\tau)\) closely matches the HPQCD prediction, and the corrected lepton-polarization observables for the \(J/\psi\) and \(\eta_c\) modes remain compatible with recent Standard Model calculations \cite{Harrison2025BcJpsiFF,Hu2019BcPQCDLat}. Thus, in the semileptonic \(S\)-wave sector, the corrected results are not only qualitatively sensible but also quantitatively close to current theoretical benchmarks.

Comparison with experiment is most meaningful through ratios. The corrected prediction for
\[
\frac{\mathcal{B}(B_c^-\to J/\psi K^-)}{\mathcal{B}(B_c^-\to J/\psi \pi^-)}
\]
is in excellent agreement with the LHCb measurement \cite{Aaij2016BcJpsiKPi}. The ratio
\[
\frac{\mathcal{B}(B_c^-\to J/\psi \pi^-)}{\mathcal{B}(B_c^-\to J/\psi \mu\bar\nu_\mu)}
\]
comes out somewhat above the LHCb value, but still of the correct order \cite{Aaij2014BcJpsiPiMuNu}. Likewise, the predicted enhancement of the \(\rho\)-dominated channel relative to \(J/\psi\pi\) is broadly consistent with the observed \(J/\psi\pi^+\pi^0\) signal once the measured \(\rho\)-dominant dynamics is taken into account \cite{Aaij2024BcJpsiPiPi0}. Finally, the present \(R(J/\psi)\) lies far below the original LHCb central value but agrees with the more recent CMS results, which are consistent with the Standard Model within uncertainties \cite{Aaij2018RJpsi,CMS2024RJpsi,CMS2025RJpsi}. Taken together, these comparisons indicate that the corrected \(S\)-wave sector compares reasonably well with both modern Standard Model theory and the currently available data.

\begin{figure}[H]
	\centering
	\begin{subfigure}{0.48\textwidth}
		\centering
		\includegraphics[width=\linewidth]{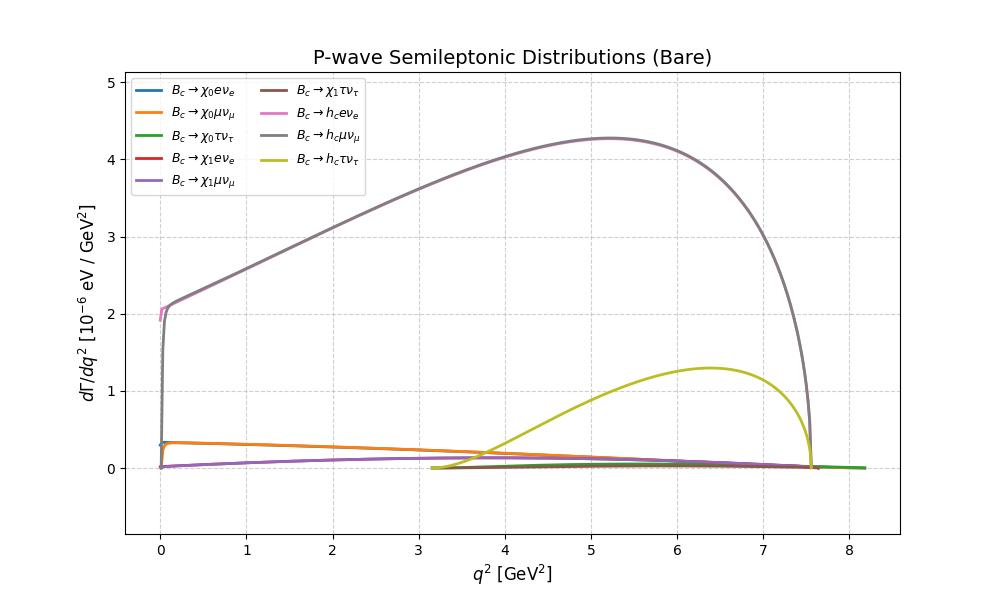}
	\end{subfigure}
	\hfill
	\begin{subfigure}{0.48\textwidth}
		\centering
		\includegraphics[width=\linewidth]{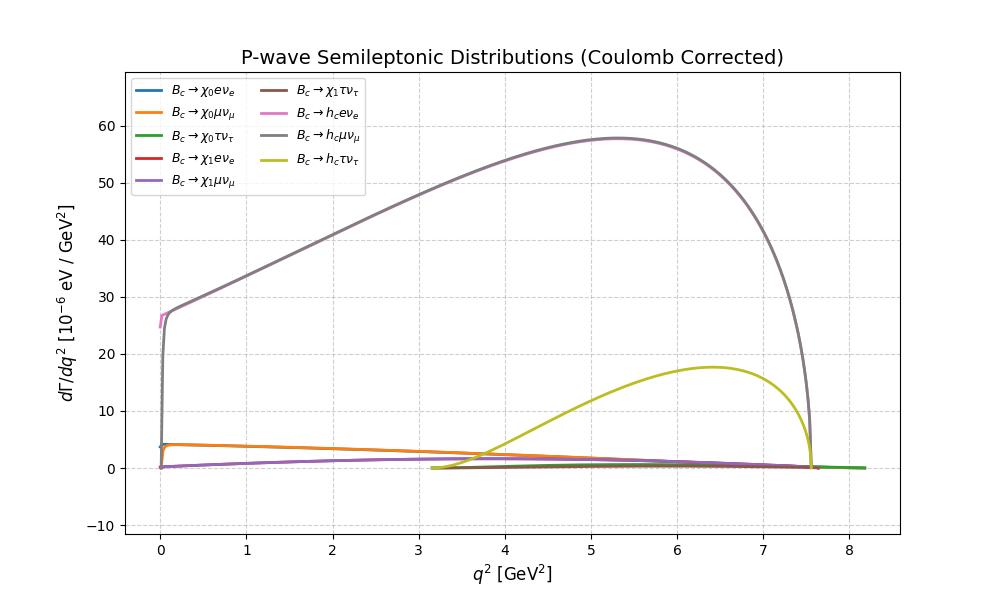}
	\end{subfigure}
	\caption{Decay-width distributions in the \(P\)-wave sector for the bare and Coulomb-corrected variants.}
	\label{fig:pwave_dist}
\end{figure}

\begin{figure}[H]
	\centering
	\begin{subfigure}{0.48\textwidth}
		\centering
		\includegraphics[width=\linewidth]{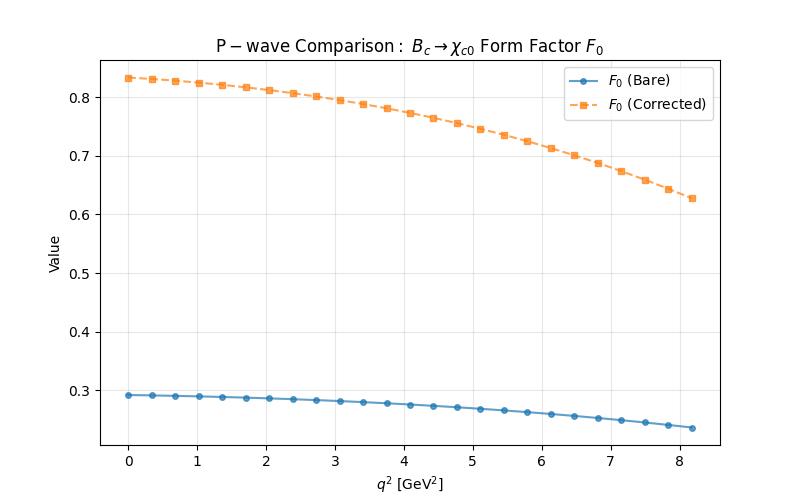}
	\end{subfigure}
	\hfill
	\begin{subfigure}{0.48\textwidth}
		\centering
		\includegraphics[width=\linewidth]{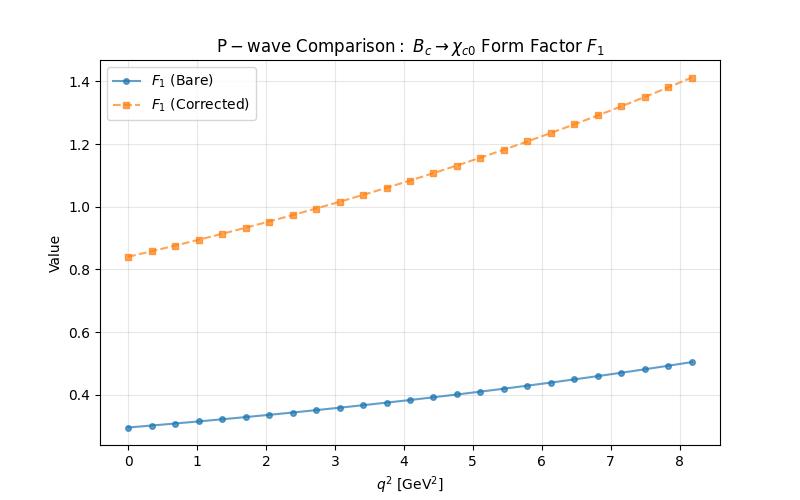}
	\end{subfigure}
	\caption{Bare and Coulomb-corrected form factors \(F_0(q^2)\) and \(F_1(q^2)\) for \(B_c\to\chi_{c0}\).}
	\label{fig:pwave_ff_chic0}
\end{figure}

\begin{figure}[H]
	\centering
	\begin{subfigure}{0.48\textwidth}
		\centering
		\includegraphics[width=\linewidth]{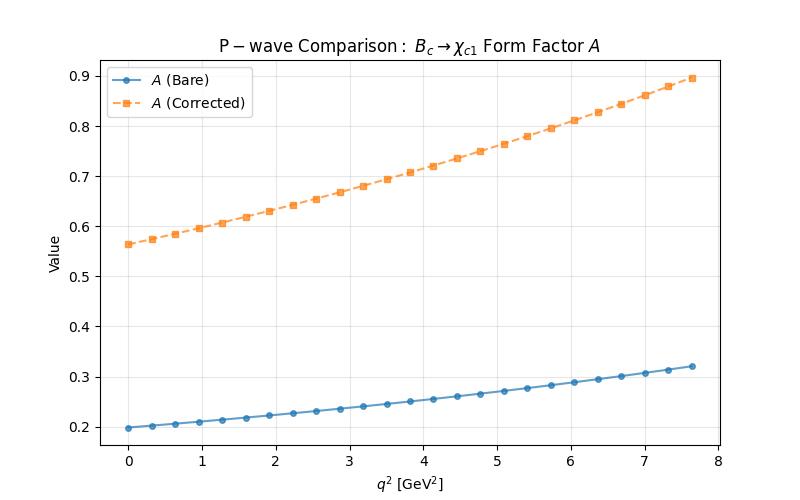}
	\end{subfigure}
	\hfill
	\begin{subfigure}{0.48\textwidth}
		\centering
		\includegraphics[width=\linewidth]{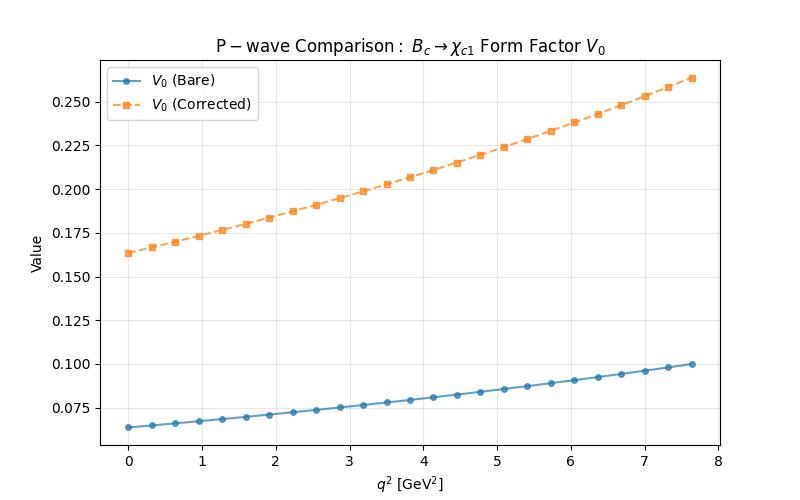}
	\end{subfigure}
	
	\vspace{0.8em}
	
	\begin{subfigure}{0.48\textwidth}
		\centering
		\includegraphics[width=\linewidth]{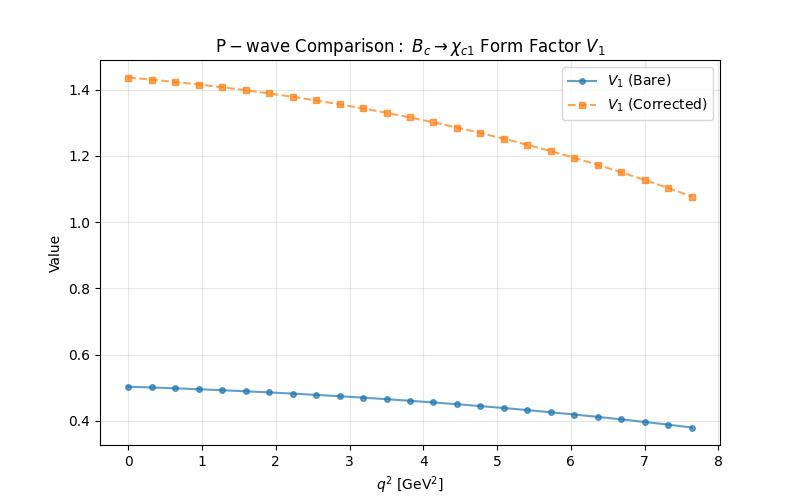}
	\end{subfigure}
	\hfill
	\begin{subfigure}{0.48\textwidth}
		\centering
		\includegraphics[width=\linewidth]{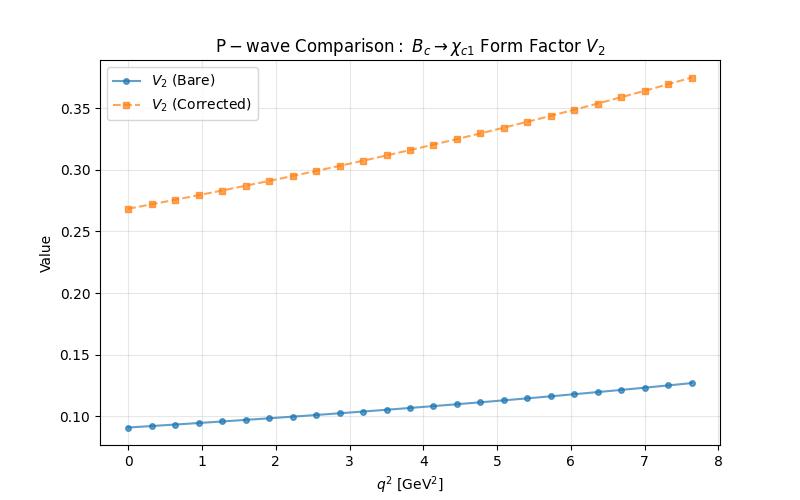}
	\end{subfigure}
	\caption{Bare and Coulomb-corrected form factors \(A(q^2)\), \(V_0(q^2)\), \(V_1(q^2)\), and \(V_2(q^2)\) for \(B_c\to\chi_{c1}\).}
	\label{fig:pwave_ff_chic1}
\end{figure}

\begin{figure}[H]
	\centering
	\begin{subfigure}{0.48\textwidth}
		\centering
		\includegraphics[width=\linewidth]{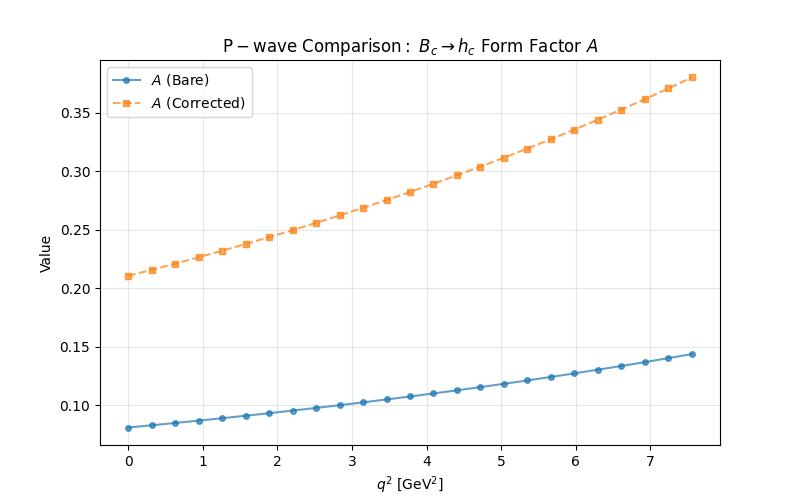}
	\end{subfigure}
	\hfill
	\begin{subfigure}{0.48\textwidth}
		\centering
		\includegraphics[width=\linewidth]{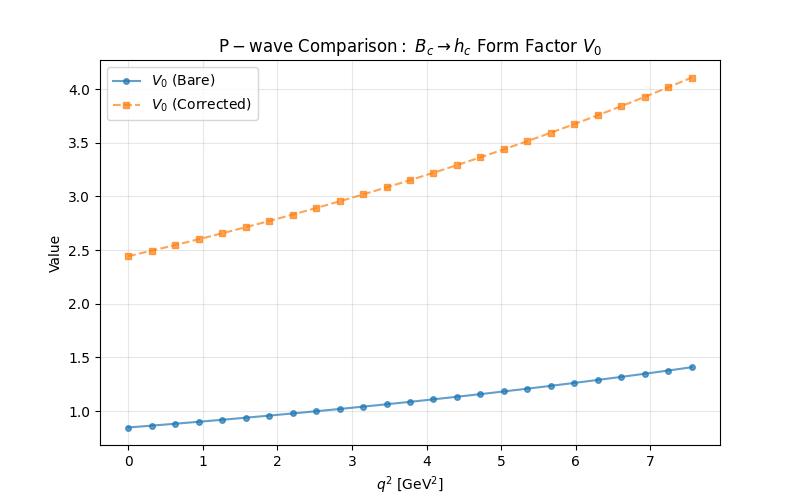}
	\end{subfigure}
	
	\vspace{0.8em}
	
	\begin{subfigure}{0.48\textwidth}
		\centering
		\includegraphics[width=\linewidth]{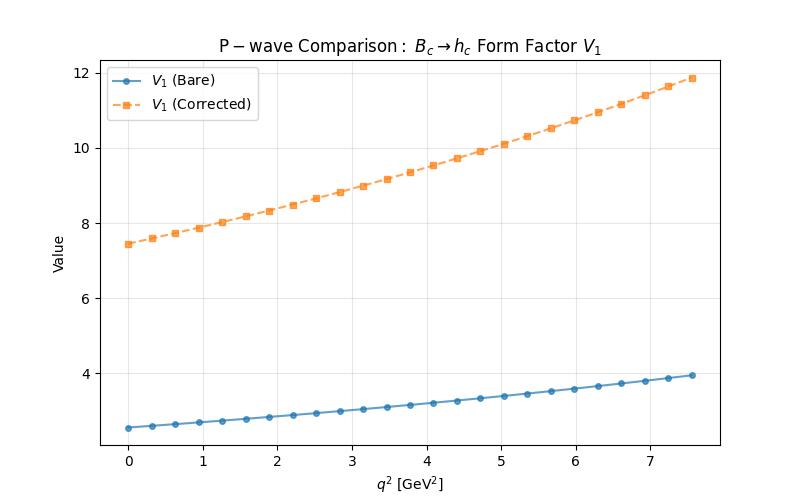}
	\end{subfigure}
	\hfill
	\begin{subfigure}{0.48\textwidth}
		\centering
		\includegraphics[width=\linewidth]{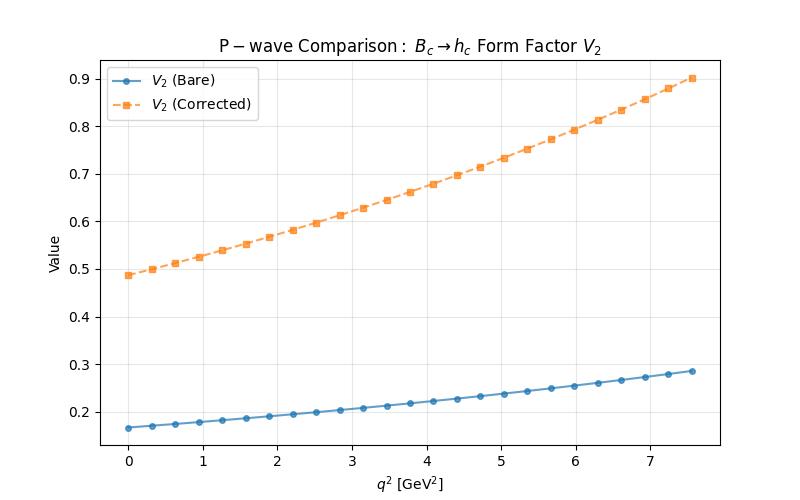}
	\end{subfigure}
	\caption{Bare and Coulomb-corrected form factors \(A(q^2)\), \(V_0(q^2)\), \(V_1(q^2)\), and \(V_2(q^2)\) for \(B_c\to h_c\).}
	\label{fig:pwave_ff_hc}
\end{figure}

The \(P\)-wave sector, summarized in Tables~\ref{tab:PWaveWidths}--\ref{tab:P_q2_hc} and Figs.~\ref{fig:pwave_dist}--\ref{fig:pwave_ff_hc}, is more differentiated among final states and therefore probes the internal spin structure of the framework more stringently. A central result is the persistent dominance of the \(h_c\) channels. Both semileptonic and nonleptonic modes involving \(h_c\) are generally larger than the corresponding \(\chi_{c0}\) and \(\chi_{c1}\) modes, often by a substantial margin. This hierarchy is consistent with recent systematic QCD sum-rule studies and is broadly compatible with earlier Bethe--Salpeter and light-cone sum-rule analyses, which also find a highly non-uniform distribution of strength across the \(P\)-wave multiplet \cite{Lu:2025bvi,Wang2011BcPWaveBS,Akan2022BcPWaveLCSR}. For nonleptonic \(P\)-wave channels, the same pattern is supported by the NLO-NRQCD analysis of \(B_c^+\to \chi_{cJ}(h_c)\pi^+\), which emphasizes the sensitivity of these two-body modes to short-distance corrections \cite{Chen2021BcPwaveNLO}.

\begin{longtable}{>{\raggedright\arraybackslash}p{5.4cm}cccc}
	\caption{\label{tab:PWaveWidths}P-wave decay widths. This work is compared with Ref.~\cite{Lu:2025bvi}. All widths in this table are in units of \(10^{-6}~\mathrm{eV}\).}\\
	\toprule
	Decay & This work (bare) & Ref.~\cite{Lu:2025bvi} (bare) & This work (corr.) & Ref.~\cite{Lu:2025bvi} (corr.)\\
	\midrule
	\endfirsthead
	\toprule
	Decay & This work (bare) & Ref.~\cite{Lu:2025bvi} (bare) & This work (corr.) & Ref.~\cite{Lu:2025bvi} (corr.)\\
	\midrule
	\endhead
	\midrule
	\multicolumn{5}{r}{Continued on next page}\\
	\endfoot
	\bottomrule
	\endlastfoot
	$\Bc^- \to \chi_{c0} e\bar\nu_e$ & 1.451 & 1.25 & 17.960 & 9.99\\
	$\Bc^- \to \chi_{c0} \mu\bar\nu_\mu$ & 1.440 & 1.24 & 17.852 & 9.93\\
	$\Bc^- \to \chi_{c0} \tau\bar\nu_\tau$ & 0.163 & 0.14 & 1.888 & 1.05\\
	$\Bc^- \to \chi_{c1} e\bar\nu_e$ & 0.708 & 0.61 & 8.606 & 4.85\\
	$\Bc^- \to \chi_{c1} \mu\bar\nu_\mu$ & 0.697 & 0.60 & 8.552 & 4.82\\
	$\Bc^- \to \chi_{c1} \tau\bar\nu_\tau$ & 0.074 & 0.064 & 0.905 & 0.51\\
	$\Bc^- \to h_c e\bar\nu_e$ & 25.718 & 22.15 & 344.047 & 194.23\\
	$\Bc^- \to h_c \mu\bar\nu_\mu$ & 25.602 & 22.05 & 342.418 & 193.31\\
	$\Bc^- \to h_c \tau\bar\nu_\tau$ & 3.518 & 3.03 & 47.684 & 26.92\\
	$\Bc^- \to \chi_{c2} K^-$ & 1.684 & 1.45 & 18.761 & 10.63\\
	$\Bc^- \to \chi_{c2} \rho^-$ & 61.189 & 52.70 & 684.472 & 387.82\\
	$\Bc^- \to \chi_{c2} K^{*-}$ & 3.251 & 2.80 & 36.375 & 20.61\\
	$\Bc^- \to \chi_{c0} \pi^-$ & 0.360 & 0.310 & 4.548 & 2.530\\
	$\Bc^- \to \chi_{c2} \pi^-$ & 2.262 & 1.948 & 25.082 & 14.211\\
	$\Bc^- \to h_c \pi^-$ & 2.800 & 2.412 & 35.368 & 19.967\\
	$\Bc^- \to \chi_{c0} K^-$ & 0.290 & 0.249 & 3.497 & 1.945\\
	$\Bc^- \to h_c K^-$ & 2.143 & 1.846 & 27.547 & 15.552\\
	$\Bc^- \to \chi_{c0} \rho^-$ & 9.545 & 8.221 & 119.919 & 66.705\\
	$\Bc^- \to \chi_{c1} \rho^-$ & 1.012 & 0.871 & 11.954 & 6.737\\
	$\Bc^- \to h_c \rho^-$ & 71.162 & 61.290 & 927.397 & 523.557\\
	$\Bc^- \to \chi_{c0} K^{*-}$ & 0.488 & 0.420 & 6.262 & 3.483\\
	$\Bc^- \to \chi_{c1} K^{*-}$ & 0.060 & 0.052 & 0.735 & 0.414\\
	$\Bc^- \to h_c K^{*-}$ & 3.989 & 3.436 & 51.585 & 29.122\\
\end{longtable}

\begin{table}[H]
	\centering
	\caption{Branching fractions for the numerically recovered \(P\)-wave channels in this work, in percent.}
	\label{tab:PWaveBR}
	\begin{tabular}{lcc}
		\toprule
		Decay channel & This work (bare) & This work (corrected) \\
		\midrule
		$B_c^- \to \chi_{c0} e\bar\nu_e$       & 0.112 & 1.392 \\
		$B_c^- \to \chi_{c0} \mu\bar\nu_\mu$   & 0.112 & 1.383 \\
		$B_c^- \to \chi_{c0} \tau\bar\nu_\tau$ & 0.013 & 0.146 \\
		$B_c^- \to \chi_{c1} e\bar\nu_e$       & 0.055 & 0.667 \\
		$B_c^- \to \chi_{c1} \mu\bar\nu_\mu$   & 0.054 & 0.663 \\
		$B_c^- \to \chi_{c1} \tau\bar\nu_\tau$ & 0.006 & 0.070 \\
		$B_c^- \to h_c e\bar\nu_e$             & 1.993 & 26.659 \\
		$B_c^- \to h_c \mu\bar\nu_\mu$         & 1.984 & 26.533 \\
		$B_c^- \to h_c \tau\bar\nu_\tau$       & 0.273 & 3.695 \\
		$B_c^- \to \chi_{c2} K^-$              & 0.130 & 1.454 \\
		$B_c^- \to \chi_{c2} \rho^-$           & 4.742 & 53.037 \\
		$B_c^- \to \chi_{c2} K^{*-}$           & 0.252 & 2.819 \\
		$B_c^- \to \chi_{c0} \pi^-$            & 0.028 & 0.352 \\
		$B_c^- \to \chi_{c2} \pi^-$            & 0.175 & 1.943 \\
		$B_c^- \to h_c \pi^-$                  & 0.217 & 2.741 \\
		$B_c^- \to \chi_{c0} K^-$              & 0.022 & 0.271 \\
		$B_c^- \to h_c K^-$                    & 0.166 & 2.135 \\
		$B_c^- \to \chi_{c0} \rho^-$           & 0.740 & 9.292 \\
		$B_c^- \to \chi_{c1} \rho^-$           & 0.078 & 0.926 \\
		$B_c^- \to h_c \rho^-$                 & 5.514 & 71.857 \\
		$B_c^- \to \chi_{c0} K^{*-}$           & 0.038 & 0.485 \\
		$B_c^- \to \chi_{c1} K^{*-}$           & 0.005 & 0.057 \\
		$B_c^- \to h_c K^{*-}$                 & 0.309 & 3.997 \\
		\bottomrule
	\end{tabular}
\end{table}

\begin{table}[H]
	\centering
	\caption{Integrated angular and polarization observables in the \(P\)-wave semileptonic sector.}
	\label{tab:PWaveIntegrated}
	\begin{tabular}{llcccccc}
		\toprule
		Channel & Mode & $\langle A_{FB}\rangle_{\rm bare}$ & $\langle A_{FB}\rangle_{\rm corr}$ & $\langle P_L^\ell\rangle_{\rm bare}$ & $\langle P_L^\ell\rangle_{\rm corr}$ & $\langle F_L\rangle_{\rm bare}$ & $\langle F_L\rangle_{\rm corr}$ \\
		\midrule
		\multirow{2}{*}{$B_c\to\chi_{c0}\ell\bar\nu_\ell$}
		& $\mu$  & 0.0166 & 0.0166 & -0.9558 & -0.9556 & --- & --- \\
		& $\tau$ & 0.4027 & 0.4036 & +0.2257 & +0.2066 & --- & --- \\
		\midrule
		\multirow{2}{*}{$B_c\to\chi_{c1}\ell\bar\nu_\ell$}
		& $\mu$  & 0.598 & 0.605 & -0.9923 & -0.9928 & 0.172 & 0.164 \\
		& $\tau$ & 0.509 & 0.508 & -0.525 & -0.530 & 0.183 & 0.179 \\
		\midrule
		\multirow{2}{*}{$B_c\to h_c\ell\bar\nu_\ell$}
		& $\mu$  & 0.0551 & 0.0495 & -0.9803 & -0.9811 & 0.537 & 0.533 \\
		& $\tau$ & 0.1676 & 0.1602 & -0.4436 & -0.4505 & 0.447 & 0.443 \\
		\bottomrule
	\end{tabular}
\end{table}

\begin{table}[H]
	\centering
	\caption{Representative corrected \(q^2\)-dependent observables for \(B_c\to \chi_{c0}\ell\bar\nu_\ell\). Differential widths are in units of \(10^{-6}\,\mathrm{eV/GeV^2}\).}
	\label{tab:P_q2_chic0}
	\begin{tabular}{llcccc}
		\toprule
		Mode & $q^2$ (GeV$^2$) & $d\Gamma/dq^2$ & $A_{FB}(q^2)$ & $P_L^\ell(q^2)$ \\
		\midrule
		\multirow{3}{*}{$\mu$}
		& 1.645 & 3.568 & 0.0100 & -0.9735 \\
		& 4.095 & 2.304 & 0.00420 & -0.9886 \\
		& 6.546 & 0.768 & 0.00325 & -0.9900 \\
		\midrule
		\multirow{3}{*}{$\tau$}
		& 4.171 & 0.341 & 0.454 & 0.231 \\
		& 5.674 & 0.609 & 0.403 & 0.150 \\
		& 7.177 & 0.339 & 0.366 & 0.269 \\
		\bottomrule
	\end{tabular}
\end{table}

\begin{table}[H]
	\centering
	\caption{Representative corrected \(q^2\)-dependent observables for \(B_c\to \chi_{c1}\ell\bar\nu_\ell\). Differential widths are in units of \(10^{-6}\,\mathrm{eV/GeV^2}\).}
	\label{tab:P_q2_chic1}
	\begin{tabular}{llccccc}
		\toprule
		Mode & $q^2$ (GeV$^2$) & $d\Gamma/dq^2$ & $A_{FB}(q^2)$ & $P_L^\ell(q^2)$ & $F_L(q^2)$ \\
		\midrule
		\multirow{3}{*}{$\mu$}
		& 1.536 & 1.080 & 0.577 & -0.9903 & 0.189 \\
		& 3.823 & 1.628 & 0.631 & -0.9967 & 0.126 \\
		& 6.110 & 1.065 & 0.627 & -0.9980 & 0.162 \\
		\midrule
		\multirow{3}{*}{$\tau$}
		& 4.062 & 0.115 & 0.500 & -0.387 & 0.157 \\
		& 5.401 & 0.312 & 0.526 & -0.515 & 0.157 \\
		& 6.741 & 0.258 & 0.509 & -0.604 & 0.207 \\
		\bottomrule
	\end{tabular}
\end{table}

\begin{table}[H]
	\centering
	\caption{Representative corrected \(q^2\)-dependent observables for \(B_c\to h_c\ell\bar\nu_\ell\). Differential widths are in units of \(10^{-6}\,\mathrm{eV/GeV^2}\).}
	\label{tab:P_q2_hc}
	\begin{tabular}{llccccc}
		\toprule
		Mode & $q^2$ (GeV$^2$) & $d\Gamma/dq^2$ & $A_{FB}(q^2)$ & $P_L^\ell(q^2)$ & $F_L(q^2)$ \\
		\midrule
		\multirow{3}{*}{$\mu$}
		& 1.520 & 37.448 & 0.0444 & -0.9768 & 0.732 \\
		& 3.784 & 52.757 & 0.0560 & -0.9941 & 0.512 \\
		& 6.048 & 55.748 & 0.0447 & -0.9976 & 0.389 \\
		\midrule
		\multirow{3}{*}{$\tau$}
		& 4.046 & 4.504  & 0.291 & -0.144 & 0.597 \\
		& 5.363 & 14.113 & 0.196 & -0.387 & 0.478 \\
		& 6.679 & 17.389 & 0.109 & -0.564 & 0.384 \\
		\bottomrule
	\end{tabular}
\end{table}

The \(\chi_{c0}\) and \(\chi_{c1}\) channels remain smaller than the \(h_c\) modes, but they still carry useful dynamical information. The \(\chi_{c0}\) mode depends on a comparatively simple form-factor structure, whereas \(B_c\to\chi_{c1}\ell\bar\nu_\ell\) is more sensitive to the competition between longitudinal and transverse helicity amplitudes. This difference is reflected directly in the integrated observables: the \(\chi_{c1}\) mode shows a large forward-backward asymmetry and a relatively small longitudinal fraction, while the \(h_c\) channel remains more strongly longitudinal. The corresponding \(q^2\)-dependent distributions show that these features arise from the evolution of the invariant amplitudes across phase space rather than from phase-space integration alone. In this structural sense, the present \(P\)-wave results are in line with modern theory.

The comparison with experiment is more demanding in the nonleptonic tensor channel. LHCb has observed \(B_c^+\to\chi_{c2}\pi^+\) and measured a branching-fraction ratio to \(J/\psi\pi^+\) well below the corresponding corrected ratio implied by the present tables \cite{Aaij2024BcChiCpi}. The present framework therefore identifies the tensor \(P\)-wave channel as phenomenologically relevant, but appears to overestimate its nonleptonic normalization. The \(P\)-wave sector should thus be regarded as qualitatively informative but quantitatively more sensitive than the \(S\)-wave sector to the treatment of factorization, form-factor normalization, and higher-order effects.

\begin{figure}[H]
	\centering
	\begin{subfigure}{0.48\textwidth}
		\centering
		\includegraphics[width=\linewidth]{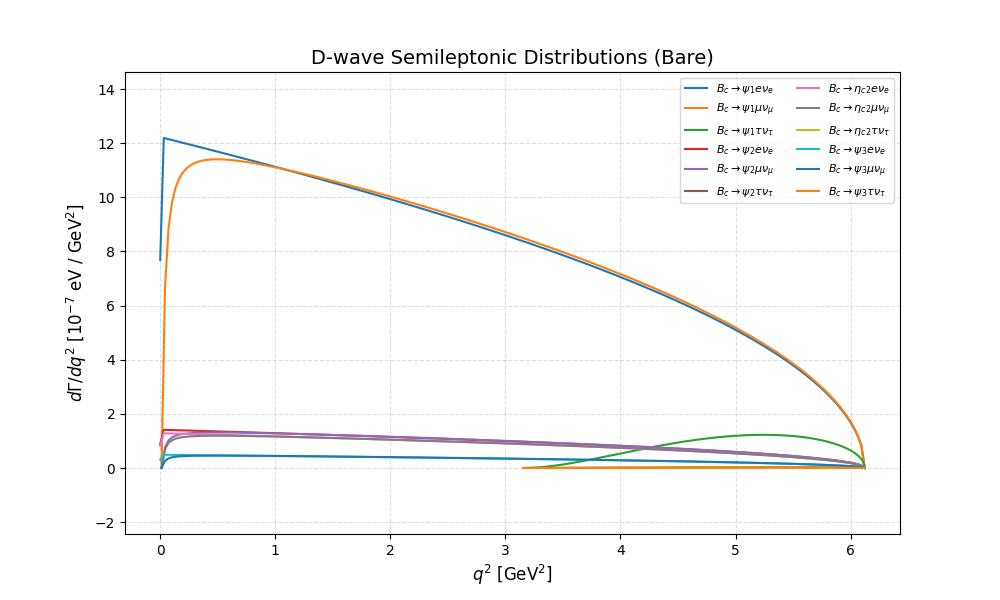}
	\end{subfigure}
	\hfill
	\begin{subfigure}{0.48\textwidth}
		\centering
		\includegraphics[width=\linewidth]{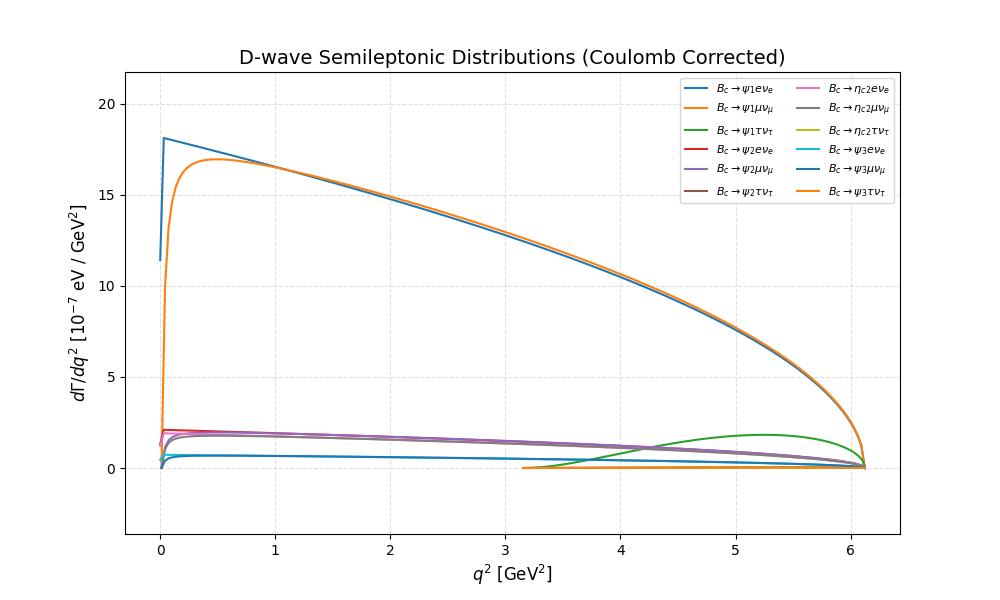}
	\end{subfigure}
	\caption{Decay-width distributions in the \(D\)-wave sector for the bare and Coulomb-corrected variants.}
	\label{fig:dwave_dist}
\end{figure}

The \(D\)-wave sector, shown in Tables~\ref{tab:DWaveWidths} and \ref{tab:DWaveBR} together with Fig.~\ref{fig:dwave_dist}, occupies the most suppressed end of the weak-decay hierarchy studied here. Even so, its internal ordering is clear and stable. The \(\psi_1\) channels dominate, followed by \(\psi_2\) and \(\eta_{c2}\), while the \(\psi_3\) modes remain smallest. This pattern agrees with the main trend emphasized in recent QCD sum-rule analyses of \(B_c\to D\)-wave charmonia \cite{Lu:2025usr} and is also compatible with the broader suppression pattern seen in earlier constituent-quark-model studies \cite{Ivanov2005BcCharmonia}. The Coulomb correction increases the overall normalization but does not reorder the multiplet, which is a nontrivial stability test of the framework.

\begin{longtable}{>{\raggedright\arraybackslash}p{5.8cm}cccc}
	\caption{\label{tab:DWaveWidths}D-wave semileptonic decay widths. This work is compared with Ref.~\cite{Lu:2025usr}. All widths in this table are in units of \(10^{-7}~\mathrm{eV}\).}\\
	\toprule
	Decay & This work (bare) & Ref.~\cite{Lu:2025usr} (bare) & This work (corr.) & Ref.~\cite{Lu:2025usr} (corr.)\\
	\midrule
	\endfirsthead
	\toprule
	Decay & This work (bare) & Ref.~\cite{Lu:2025usr} (bare) & This work (corr.) & Ref.~\cite{Lu:2025usr} (corr.)\\
	\midrule
	\endhead
	\midrule
	\multicolumn{5}{r}{Continued on next page}\\
	\endfoot
	\bottomrule
	\endlastfoot
	$\Bc^- \to \psione e^- \bar\nu_e$   & 33.3228   & 28.70   & 49.9835   & 28.70 \\
	$\Bc^- \to \psione \mu^- \bar\nu_\mu$ & 32.9861   & 28.41   & 49.4784   & 28.41 \\
	$\Bc^- \to \psione \tau^- \bar\nu_\tau$ & 1.52101   & 1.31    & 2.28148   & 1.31 \\
	$\Bc^- \to \psitwo e^- \bar\nu_e$   & 3.85477   & 3.32    & 5.73893   & 3.32 \\
	$\Bc^- \to \psitwo \mu^- \bar\nu_\mu$ & 3.79671   & 3.27    & 5.65250   & 3.27 \\
	$\Bc^- \to \psitwo \tau^- \bar\nu_\tau$ & 0.0452819 & 0.039   & 0.0674151 & 0.039 \\
	$\Bc^- \to \etatwo e^- \bar\nu_e$   & 3.50644   & 3.02    & 5.28429   & 3.02 \\
	$\Bc^- \to \etatwo \mu^- \bar\nu_\mu$ & 3.44839   & 2.97    & 5.19681   & 2.97 \\
	$\Bc^- \to \etatwo \tau^- \bar\nu_\tau$ & 0.0313490 & 0.027   & 0.0472437 & 0.027 \\
	$\Bc^- \to \psithree e^- \bar\nu_e$ & 1.33524   & 1.15    & 1.95269   & 1.15 \\
	$\Bc^- \to \psithree \mu^- \bar\nu_\mu$ & 1.32362   & 1.14    & 1.93571   & 1.14 \\
	$\Bc^- \to \psithree \tau^- \bar\nu_\tau$ & 0.0113785 & 0.0098  & 0.0166403 & 0.0098 \\
\end{longtable}

\begin{table}[H]
	\centering
	\caption{D-wave semileptonic branching fractions in this work, in units of \(10^{-3}\).}
	\label{tab:DWaveBR}
	\begin{tabular}{lcc}
		\toprule
		Decay channel & This work (bare) & This work (corrected) \\
		\midrule
		$B_c^- \to \psi_1 e^- \bar\nu_e$          & 2.582 & 3.873 \\
		$B_c^- \to \psi_1 \mu^- \bar\nu_\mu$      & 2.556 & 3.834 \\
		$B_c^- \to \psi_1 \tau^- \bar\nu_\tau$    & 0.118 & 0.177 \\
		$B_c^- \to \psi_2 e^- \bar\nu_e$          & 0.298 & 0.445 \\
		$B_c^- \to \psi_2 \mu^- \bar\nu_\mu$      & 0.294 & 0.438 \\
		$B_c^- \to \psi_2 \tau^- \bar\nu_\tau$    & 0.00349 & 0.00522 \\
		$B_c^- \to \eta_{c2} e^- \bar\nu_e$       & 0.272 & 0.409 \\
		$B_c^- \to \eta_{c2} \mu^- \bar\nu_\mu$   & 0.267 & 0.403 \\
		$B_c^- \to \eta_{c2} \tau^- \bar\nu_\tau$ & 0.00240 & 0.00366 \\
		$B_c^- \to \psi_3 e^- \bar\nu_e$          & 0.104 & 0.151 \\
		$B_c^- \to \psi_3 \mu^- \bar\nu_\mu$      & 0.102 & 0.150 \\
		$B_c^- \to \psi_3 \tau^- \bar\nu_\tau$    & 0.00088 & 0.00129 \\
		\bottomrule
	\end{tabular}
\end{table}

The \(D\)-wave branching fractions and semileptonic ratios are especially useful because they partly factor out common normalizations and expose the internal hierarchy more directly. They also show that \(\tau\)-mode suppression becomes increasingly severe from the dominant \(S\)-wave channels to the higher-spin \(D\)-wave states, as expected from the combined phase-space and helicity suppression. Since no direct experimental information on exclusive \(B_c\) decays to \(D\)-wave charmonia is presently available, this sector should be viewed mainly as a prediction of the model rather than as a calibrated part of it.

Taken together, the weak-decay analysis gives a coherent picture across all orbital sectors, but the quality of external validation is not uniform. The corrected \(S\)-wave sector compares rather well with both modern Standard Model calculations and the currently available experimental ratios. The \(P\)-wave sector reproduces the qualitative hierarchy found in modern theory and correctly identifies tensor channels as relevant, but its nonleptonic normalization appears less robust. The \(D\)-wave sector remains theoretically well ordered and consistent with recent sum-rule expectations, but still awaits direct experimental tests. In this sense, the present framework is most successful as a unified structural description of the weak \(B_c\to(c\bar c)\) system, with the \(S\)-wave modes already anchored by both theory and experiment, the \(P\)-wave channels providing a more demanding stress test, and the \(D\)-wave modes serving as predictions for future measurements.

\subsection{Nonleptonic channels with charmonium plus \(D^{(*)}_{(s)}\)}

The channels containing charmonium together with \(D\), \(D^\ast\), \(D_s\), or \(D_s^\ast\) mesons form an intermediate sector between the purely charmonium modes and the broader open-charm decay pattern. Since the amplitudes combine a \(B_c\to(c\bar c)\) transition form factor with a heavy-meson decay constant, they test whether the channel dependence already seen in the semileptonic sector survives when the second final-state hadron is also heavy. The relevant comparison set includes both the recent sum-rule analysis of these modes and earlier relativistic quark-model studies, as well as the measured \(J/\psi D_s^{(*)}\) channels \cite{Ebert2003BcCharmD,Wu:2024dzn,Zhu2017BcJpsiDs,LHCb2013BcJpsiDs,ATLAS2022BcJpsiDs}.

\begin{longtable}{>{\raggedright\arraybackslash}p{6.4cm}cc}
	\caption{\label{tab:CharDWidths}Nonleptonic decay widths to charmonium plus \(D^{(*)}_{(s)}\). This work is compared with Ref.~\cite{Wu:2024dzn}. All widths are in units of \(10^{-7}~\mathrm{eV}\).}\\
	\toprule
	Decay & This work & Ref.~\cite{Wu:2024dzn}\\
	\midrule
	\endfirsthead
	\toprule
	Decay & This work & Ref.~\cite{Wu:2024dzn}\\
	\midrule
	\endhead
	\midrule
	\multicolumn{3}{r}{Continued on next page}\\
	\endfoot
	\bottomrule
	\endlastfoot
	$\Bc^- \to D^- \etac$ & 1.191 & 0.63\\
	$\Bc^- \to D^- \psij$ & 0.550 & 0.46\\
	$\Bc^- \to D^{*-}\etac$ & 1.547 & 0.86\\
	$\Bc^- \to D^{*-}\psij$ & 2.555 & 2.2\\
	$\Bc^- \to D_s^- \etac$ & 30.156 & 16\\
	$\Bc^- \to D_s^- \psij$ & 13.163 & 11\\
	$\Bc^- \to D_s^{*-}\etac$ & 37.873 & 21\\
	$\Bc^- \to D_s^{*-}\psij$ & 68.525 & 59\\
\end{longtable}

\begin{table}[H]
	\centering
	\caption{Branching fractions for the charmonium plus \(D^{(*)}_{(s)}\) nonleptonic sector in this work, in units of \(10^{-3}\).}
	\label{tab:CharD_BR}
	\begin{tabular}{lc}
		\toprule
		Decay channel & This work \\
		\midrule
		$B_c^- \to D^- \eta_c$         & 0.092 \\
		$B_c^- \to D^- J/\psi$         & 0.043 \\
		$B_c^- \to D^{*-}\eta_c$       & 0.120 \\
		$B_c^- \to D^{*-}J/\psi$       & 0.198 \\
		$B_c^- \to D_s^- \eta_c$       & 2.337 \\
		$B_c^- \to D_s^- J/\psi$       & 1.020 \\
		$B_c^- \to D_s^{*-}\eta_c$     & 2.934 \\
		$B_c^- \to D_s^{*-}J/\psi$     & 5.309 \\
		\bottomrule
	\end{tabular}
\end{table}

The widths and branching fractions listed in Tables~\ref{tab:CharDWidths} and \ref{tab:CharD_BR} show a pronounced enhancement of strange final states relative to the corresponding non-strange ones. This is exactly the hierarchy expected from the larger CKM weight of the \(c\to s\) transition and from the larger decay constants of the strange charmed mesons. The pattern agrees with both the recent sum-rule analysis of Ref.~\cite{Wu:2024dzn} and the earlier relativistic quark-model study of Ref.~\cite{Ebert2003BcCharmD}. Quantitatively, however, the present widths are systematically larger than those of Ref.~\cite{Wu:2024dzn}, with the enhancement being especially strong for the channels containing \(\eta_c\).

A second notable feature is the non-uniform spin dependence inside the \(D^{(*)}_{(s)}\) sector. For pseudoscalar open-charm mesons, the \(\eta_c\) channels remain larger than the corresponding \(J/\psi\) channels. Once the open-charm meson is promoted to a vector state, however, the pattern changes: \(D_s^{*-}J/\psi\) becomes the dominant channel of the sector, and even \(D^{*-}J/\psi\) overtakes \(D^{*-}\eta_c\). This shows that the spin sensitivity visible in the semileptonic \(B_c\to\eta_c\) and \(B_c\to J/\psi\) transitions is not washed out by factorization, but reappears in a channel-dependent way in the heavy-heavy nonleptonic sector.

The most direct experimental tests currently available are the \(J/\psi D_s^+\) and \(J/\psi D_s^{*+}\) modes measured by LHCb and ATLAS \cite{LHCb2013BcJpsiDs,ATLAS2022BcJpsiDs}. The present calculation reproduces the qualitative ordering
\[
J/\psi D_s^{*} > J/\psi D_s > J/\psi D^{*} > J/\psi D,
\]
and therefore captures the dominant flavor and spin hierarchy. However, when the strange channels are normalized to the experimentally accessible \(J/\psi\pi\) mode, the predicted \(J/\psi D_s\) and \(J/\psi D_s^\ast\) ratios are smaller than the observed values, while the internal ratio \(J/\psi D_s^\ast/J/\psi D_s\) is larger than measured. The present factorized implementation therefore reproduces the ordering of the sector more convincingly than its absolute normalization.

This conclusion is also consistent with the broader theory literature. Earlier relativistic quark-model work already found strong dominance of the strange channels, while more recent pQCD/NRQCD-improved analyses of the \(J/\psi D_s^{(*)}\) system obtained ratios closer to the measured values \cite{Ebert2003BcCharmD,Zhu2017BcJpsiDs}. The present results thus fit into a recognizable pattern: a factorized treatment is sufficient to recover the main flavor hierarchy and the broad vector-versus-pseudoscalar ordering, but the absolute normalization of the experimentally accessible strange channels remains sensitive to the detailed treatment of form factors, effective coefficients, and possible nonfactorizable QCD effects.

Overall, this sector confirms that the channel dependence of the weak amplitudes is not confined to semileptonic observables but propagates consistently into two-body hadronic decays with a second heavy meson. At the same time, it exposes one of the clearer quantitative tensions of the present framework. The strange channels are correctly identified as dominant, and the \(D_s^{*-}J/\psi\) mode naturally emerges as the leading decay of the sector, but comparison with the available \(J/\psi D_s^{(*)}\) data shows that further improvement is needed in the normalization of the factorized amplitudes.

\subsection{Purely leptonic widths, annihilation contribution, and lifetime}

The purely leptonic widths and the annihilation contribution provide a compact test of the short-distance normalization used throughout the weak sector. Unlike the exclusive semileptonic and nonleptonic modes, these observables depend directly on the decay constants and on the inclusive weak-decay bookkeeping of the \(B_c\) meson. They therefore test whether the same short-distance input that describes the spectrum and exclusive amplitudes also leads to a phenomenologically acceptable inclusive picture \cite{Godfrey2004,Gershtein1995PhysUsp,AbdElHadyMunozVary2000,Aaij2014BcLifetime,PDG2024BflavoredHadrons,RaiVinodkumar2006,Yang:2021crs,Aebischer2021BcLifetime,Aaij2015BcLifetimeJpsiPi}.

\begin{table}[H]
	\begin{center}
		\caption{\label{Table:Br}Leptonic branching fractions of the \(B_c^{+}\) and \(B_c^{*+}\) mesons. Representative comparisons are shown with Gershtein et al.~\cite{Gershtein1995PhysUsp} and Yang et al.~\cite{Yang:2021crs}.}
		\begin{tabular}{ccccc}
			\noalign{\smallskip}\hline\noalign{\smallskip}
			\noalign{\smallskip}\hline\noalign{\smallskip}
			Meson & State & $B_c^{+}\rightarrow\tau^{+}\nu_{\tau}$ & $B_c^{+}\rightarrow\mu^{+}\nu_{\mu}$ & $B_c^{+}\rightarrow e^{+}\nu_{e}$ \\
			&& $Br_{\tau}\times 10^{-6}$ & $Br_{\mu}\times 10^{-6}$ & $Br_{e}\times 10^{-6}$ \\
			\noalign{\smallskip}\hline\noalign{\smallskip}
			$B_c$ & This work & 3.3 & 1.83 & 4.3 \\
			&~\cite{Gershtein1995PhysUsp} & 1.0 & 4.7 & 4.7 \\
			\noalign{\smallskip}\hline\noalign{\smallskip}
			$B_c^{*}$ & This work & 3.18 & 3.61 & 3.61 \\
			& ~\cite{Yang:2021crs} & $3.3^{+0.4}_{-0.3}$ & $3.8^{+0.4}_{-0.3}$ & $3.8^{+0.4}_{-0.3}$ \\
			\noalign{\smallskip}\hline\noalign{\smallskip}
		\end{tabular}
	\end{center}
\end{table}

\begin{table}[H]
	\begin{center}
		\caption{\label{Table:decaywidth}Decay widths (in \(10^{-4}\,\mathrm{eV}\)) and lifetime \(\tau\) (in ps) of the \(B_c^{+}\) meson. Representative comparisons are shown with Rai and Vinodkumar~\cite{RaiVinodkumar2006}, Abd El-Hady et al.~\cite{AbdElHadyMunozVary2000}, and Godfrey~\cite{Godfrey2004}.}
		\begin{tabular}{lccc}
			\noalign{\smallskip}\hline\noalign{\smallskip}
			\noalign{\smallskip}\hline\noalign{\smallskip}
			Model & $\Gamma(\mathrm{Anni})$ & $\Gamma(B_c^{+}\rightarrow X)$ & $\tau$ (ps) \\
			\noalign{\smallskip}\hline\noalign{\smallskip}
			This work & 0.403 & 13.21 & 0.774 \\
			\cite{RaiVinodkumar2006} & 0.642 & 13.46 & 0.489 \\
			\cite{AbdElHadyMunozVary2000} & 1.40 & 14.00 & 0.47 \\
			\cite{Godfrey2004} & 0.67 & 8.8 & 0.75 \\
			\noalign{\smallskip}\hline\noalign{\smallskip}
		\end{tabular}
	\end{center}
\end{table}

The leptonic branching fractions listed in Table~\ref{Table:Br} are naturally very small and remain primarily theoretical benchmark observables. Even so, they are useful because they directly probe the decay-constant normalization. In particular, the \(B_c^{*+}\) leptonic branching fractions are close to the recent estimates of Yang \textit{et al.}~\cite{Yang:2021crs}, indicating that the vector decay constant is of the expected magnitude. The pseudoscalar \(B_c^+\) leptonic modes likewise remain within the broad historical range of model predictions represented by Gershtein \textit{et al.}~\cite{Gershtein1995PhysUsp}, so at the level of overall scale the weak-annihilation normalization is not anomalous.

The annihilation contribution in Table~\ref{Table:decaywidth} is clearly subleading relative to the total inclusive weak width, but it is not negligible. This is the standard picture of the \(B_c\) meson, in which the dominant contribution to the full width is generated by spectator decays of the heavy constituents, while annihilation provides a smaller correction \cite{Aebischer2021BcLifetime,Beneke1996BcLifetime}. From that perspective, the present value of \(\Gamma(\mathrm{Anni})\) is of the expected order: it is well below the total width but still large enough to contribute visibly to the inclusive decay budget.

The main tension appears in the lifetime. The predicted value remains within the broad range of older model calculations and is numerically close to some earlier estimates, but it is longer than the current experimental world average and the precise LHCb determinations in the \(J/\psi\mu\nu\) and \(J/\psi\pi\) channels \cite{Aaij2014BcLifetime,PDG2024BflavoredHadrons,Aaij2015BcLifetimeJpsiPi}. Modern Standard Model OPE analyses also favor the shorter experimental scale within uncertainties \cite{Aebischer2021BcLifetime}. The appropriate conclusion is therefore that the present inclusive treatment reproduces the expected hierarchy of contributions, but still underestimates the total weak width.

This point is important for the global interpretation of the model. The same short-distance normalization that performs reasonably well in several exclusive channels leads to a plausible annihilation sector, but the inclusive lifetime shows that the total decay rate is still not fully satisfactory. The purely leptonic and annihilation observables should therefore be viewed as a useful stress test of the framework: they support its basic internal consistency, while at the same time indicating where further refinement of the inclusive weak-decay treatment is needed.

\subsection{Radiative transitions}

The radiative sector probes overlaps between different radial and orbital wave functions and therefore provides a complementary test of the spectroscopic framework. The electric-dipole transitions are mainly sensitive to orbital structure, whereas the magnetic-dipole transitions probe spin-flip effects and short-distance radial overlaps. The results are summarized in Tables~\ref{Table:bcE1} and \ref{Table:bcm1}.

\begin{table}[H]
	\begin{center}
		\caption{\label{Table:bcE1}\(E1\) transition widths of the \(B_c\) meson. Results of this work are compared with representative quark-model calculations \cite{EichtenQuigg1994,Gershtein1995PRD,Ebert2003Prop,DevlaniRai2014,Monteiro2017}.}
		\resizebox{\textwidth}{!}{
			\begin{tabular}{c c c c c c c c c c}
				\noalign{\smallskip}\hline\noalign{\smallskip} 
				\noalign{\smallskip}\hline\noalign{\smallskip}
				Initial & Final & $E_{\gamma}$ (MeV) & This work (keV) & ~\cite{DevlaniRai2014} & ~\cite{Ebert2003Prop} &~\cite{EichtenQuigg1994} &~\cite{Gershtein1995PRD} & ~\cite{Fulcher1999Bc} &~\cite{Monteiro2017} \\	\noalign{\smallskip}\hline\noalign{\smallskip}
				$B_{c}(1^3P_{2})$&$B_{c}(1^3S_{1})$&444&82.79&64.24&107&112.6&102.4&126&89.04\\
				$B_{c}(1^{1}P_{1})$&$B_{c}(1^3S_{1})$&444&9.65&9.98&13.6&0.1&8.1&26.2&121.14\\
				$B_{c}(1^{1}P_{1})$&$B_{c}(1^1S_{0})$&530&135.48&72.28&132&56.4&131.1&128&148.99\\
				$B_{c}(1^{3}P_{1})$&$B_{c}(1^3S_{1})$&443&79.15&71.32&78.9&99.5&77.8&75.8&83.88\\
				$B_{c}(1^{3}P_{1})$&$B_{c}(1^1S_{0})$&52.9&13.77&9.83&18.4&0.01&1.6&32.5&106.09\\
				$B_{c}(1^3P_{0})$&$B_{c}(1^3S_{1})$&417&65.92&58.56&67.2&79.2&65.3&74.2&42.38\\
				\noalign{\smallskip}\hline\noalign{\smallskip}
				$B_{c}(2^3S_{1})$&$B_{c}(1^3P_{2})$&0.44&2.08&2.28&5.18&17.7&14.8&14.5&0.269\\
				$B_{c}(2^3S_{1})$&$B_{c}(1^{1}P_{1})$&0.507&0.25&0.26&0.63&0.0&1.0&2.5&0.00751\\
				$B_{c}(2^3S_{1})$&$B_{c}(1^{3}P_{1})$&0.577&1.06&1.45&5.05&14.5&12.8&13.3&0.384\\
				$B_{c}(2^3S_{1})$&$B_{c}(1^3P_{0})$&0.854&0.115&0.94&3.78&7.8&7.7&9.6&3.684\\
				$B_{c}(2^1S_{0})$&$B_{c}(1^{1}P_{1})$&0.448&5.01&3.03&3.72&5.2&15.9&13.1&0.00068\\
				$B_{c}(2^1S_{0})$&$B_{c}(1^{3}P_{1})$&0.458&53.5&73.98&1.02&0.0&1.9&6.4&0.238\\
				\noalign{\smallskip}\hline\noalign{\smallskip}
				$B_{c}(1^3D_{3})$&$B_{c}(1^3P_{2})$&228&6.2&20.57&5.52&2.7&2.2&&\\
				$B_{c}(1^{1}D_{2})$&$B_{c}(1^3P_{2})$&228&11.49&18.17&12.8&&6.8&&\\
				$B_{c}(1^{3}D_{2})$&$B_{c}(1^3P_{2})$&224&26.47&2.15&27.5&24.7&12.2&&\\
				$B_{c}(1^3D_{1})$&$B_{c}(1^3P_{2})$&228&3.09&5.55&102&98.7&76.9&&\\
				$B_{c}(1^3D_{1})$&$B_{c}(1^{1}P_{1})$&239&53.90&29.61&7.66&0.0&3.3&&\\
				$B_{c}(1^3D_{1})$&$B_{c}(1^{3}P_{1})$&240&54.55&57.76&73.8&49.3&39.2&&\\
				$B_{c}(1^3D_{1})$&$B_{c}(1^3P_{0})$&267&100.034&65.75&128&88.6&79.7&&\\
				$B_{c}(1^{1}D_{2})$&$B_{c}(1^{1}P_{1})$&240&8.205&1.15&116&92.5&46.6&&\\
				$B_{c}(1^{1}D_{2})$&$B_{c}(1^{3}P_{1})$&241&0.99&0.99&7.25&&25.0&&\\
				$B_{c}(1^{3}D_{2})$&$B_{c}(1^{1}P_{1})$&235&92.41&79.67&14.1&0.1&15.4&&\\
				$B_{c}(1^{3}D_{2})$&$B_{c}(1^{3}P_{1})$&240&98.20&64.92&112&88.8&45.6&&\\
				\noalign{\smallskip}\hline\noalign{\smallskip}
				$B_{c}(2^3P_{2})$&$B_{c}(2^3S_{1})$&284&89.79&15.11&57.3&73.8&49.4&&85.639\\
				$B_{c}(2^{1}P_{1})$&$B_{c}(2^3S_{1})$&275&81.84&56.28&9.07&5.4&5.9&&158.596\\
				$B_{c}(2^{1}P_{1})$&$B_{c}(2^1S_{0})$&286&92.54&50.40&72.5&&58.0&&171.244\\
				$B_{c}(2^{3}P_{1})$&$B_{c}(2^3S_{1})$&275&81.84&16.52&37.9&54.3&32.1&&104.751\\
				$B_{c}(2^{3}P_{1})$&$B_{c}(2^1S_{0})$&286&92.54&55.05&11.7&&8.1&&114.223\\
				$B_{c}(2^3P_{0})$&$B_{c}(2^3S_{1})$&257&66.59&&29.2&41.2&25.5&&0.422\\
				\noalign{\smallskip}\hline\noalign{\smallskip}
				$B_{c}(2^3P_{2})$&$B_{c}(1^3D_{3})$&102&5.50&6.28&1.50&17.8&10.9&&\\
				$B_{c}(2^3P_{2})$&$B_{c}(1^{1}D_{2})$&101&0.95&0.91&0.113&&0.5&&\\
				$B_{c}(2^3P_{2})$&$B_{c}(1^{3}D_{2})$&106&1.10&1.16&0.26&3.2&1.5&&\\
				$B_{c}(2^3P_{2})$&$B_{c}(1^3D_{1})$&102&0.0655&0.07&0.035&0.2&0.1&&\\
				$B_{c}(2^{1}P_{1})$&$B_{c}(1^3D_{1})$&93&1.24&0.35&0.073&0.4&0.3&&\\
				$B_{c}(2^{3}P_{1})$&$B_{c}(1^3D_{1})$&93&1.24&1.14&0.180&0.3&1.6&&\\
				$B_{c}(2^3P_{0})$&$B_{c}(1^3D_{1})$&74&2.54&3.94&0.036&0.9&3.2&&\\
				\noalign{\smallskip}\hline\noalign{\smallskip}
			\end{tabular}
		}
	\end{center}
\end{table}

\begin{table}[H]
	\begin{center}
		\caption{\label{Table:bcm1}\(M1\) transition widths of the \(B_c\) meson. Results of this work are compared with representative quark-model calculations \cite{EichtenQuigg1994,Gershtein1995PRD,Ebert2003Prop,DevlaniRai2014,Monteiro2017}.}
		\resizebox{\textwidth}{!}{
			\begin{tabular}{c c c c c c c c c}
				\noalign{\smallskip}\hline\noalign{\smallskip}
				\noalign{\smallskip}\hline\noalign{\smallskip}
				Initial $\rightarrow$ Final & $E_{\gamma}$ (MeV) & This work (keV) & ~\cite{DevlaniRai2014} & ~\cite{Ebert2003Prop} & ~\cite{EichtenQuigg1994} & ~\cite{Gershtein1995PRD} & ~\cite{Fulcher1999Bc} &~\cite{Monteiro2017} \\	\noalign{\smallskip}\hline\noalign{\smallskip}
				$B_{c}(1^3S_{1}) \rightarrow B_{c}(1^1S_{0})$&13.9&0.0632&0.0469&0.033&0.135&0.066&0.059&0.185\\
				$B_{c}(2^3S_{1}) \rightarrow B_{c}(2^1S_{0})$&19&0.00474&0.0039&0.029&0.029&0.010&0.012&0.0018\\
				$B_{c}(3^3S_{1}) \rightarrow B_{c}(3^1S_{0})$&49&0.00075&0.0019&&&&&\\
				$B_{c}(2^3S_{1}) \rightarrow B_{c}(1^1S_{0})$&582&0.32&0.516&0.428&0.123&0.098&0.122&0.193\\
				$B_{c}(2^1S_{0}) \rightarrow B_{c}(1^3S_{1})$&486&9.92&1.124&0.488&0.093&0.096&0.139&\\
				\noalign{\smallskip}\hline\noalign{\smallskip}
				$B_{c}(1P_{1}) \rightarrow B_{c}(1^3P_{0})$&27.9&0.00111&0.0185&&1&&&\\
				$B_{c}(1^{1}P_{1}) \rightarrow B_{c}(1^3P_{0})$&0.289&0.0012&0.0091&&&&&\\
				$B_{c}(1^{3}P_{2}) \rightarrow B_{c}(1P_{1})$&0.129&0.0033&0.0005&&&&&\\
				$B_{c}(1^{1}P_{1}) \rightarrow B_{c}(1^3P_{2})$&0.120&0.0044&0.0002&&&&&\\
				\noalign{\smallskip}\hline\noalign{\smallskip}
			\end{tabular}
		}
	\end{center}
\end{table}

The \(E1\) widths display the expected pattern. The dominant transitions are concentrated in the low-lying \(P\to S\) sector and in selected \(D\to P\) and \(2P\to2S\) cascades, where both the dipole matrix element and the photon energy are favorable. The calculated widths fall in the same broad tens-to-hundreds of keV range found in representative quark-model studies \cite{EichtenQuigg1994,Godfrey2004,Ebert2003Prop,DevlaniRai2014,Monteiro2017}. The agreement is not exact channel by channel, but the overall scale and ordering are consistent with the wider literature.

The \(M1\) widths are much smaller, especially for hindered transitions. This is expected, because once radial orthogonality suppresses the leading overlap, relatively small changes in the wave functions or photon energies can generate large fractional variations in the widths. The allowed \(1^3S_1\to1^1S_0\) transition remains within the standard band of quark-model results, which supports the short-distance spin structure of the present wave functions. Overall, the radiative sector provides a useful complementary validation of the spectroscopic input, confirming the established hierarchy of dominant \(E1\) and suppressed \(M1\) transitions.

\subsection{Regge trajectories and global spectral organization}

The Regge analysis provides a compact global summary of the \(B_c\) spectrum by reducing the state-by-state mass information to approximately linear relations in the \((J,M^2)\) and \((n_r,M^2)\) planes. Its purpose is not to replace the underlying Hamiltonian calculation, but to test whether the predicted levels exhibit the same broad geometric regularity found in earlier studies of heavy quarkonia and of the \(B_c\) system in particular. In that sense, it serves as a useful global consistency check on the screened-potential spectrum.

\begin{figure*}[!htb]
	\centering
	\includegraphics[width=0.48\textwidth]{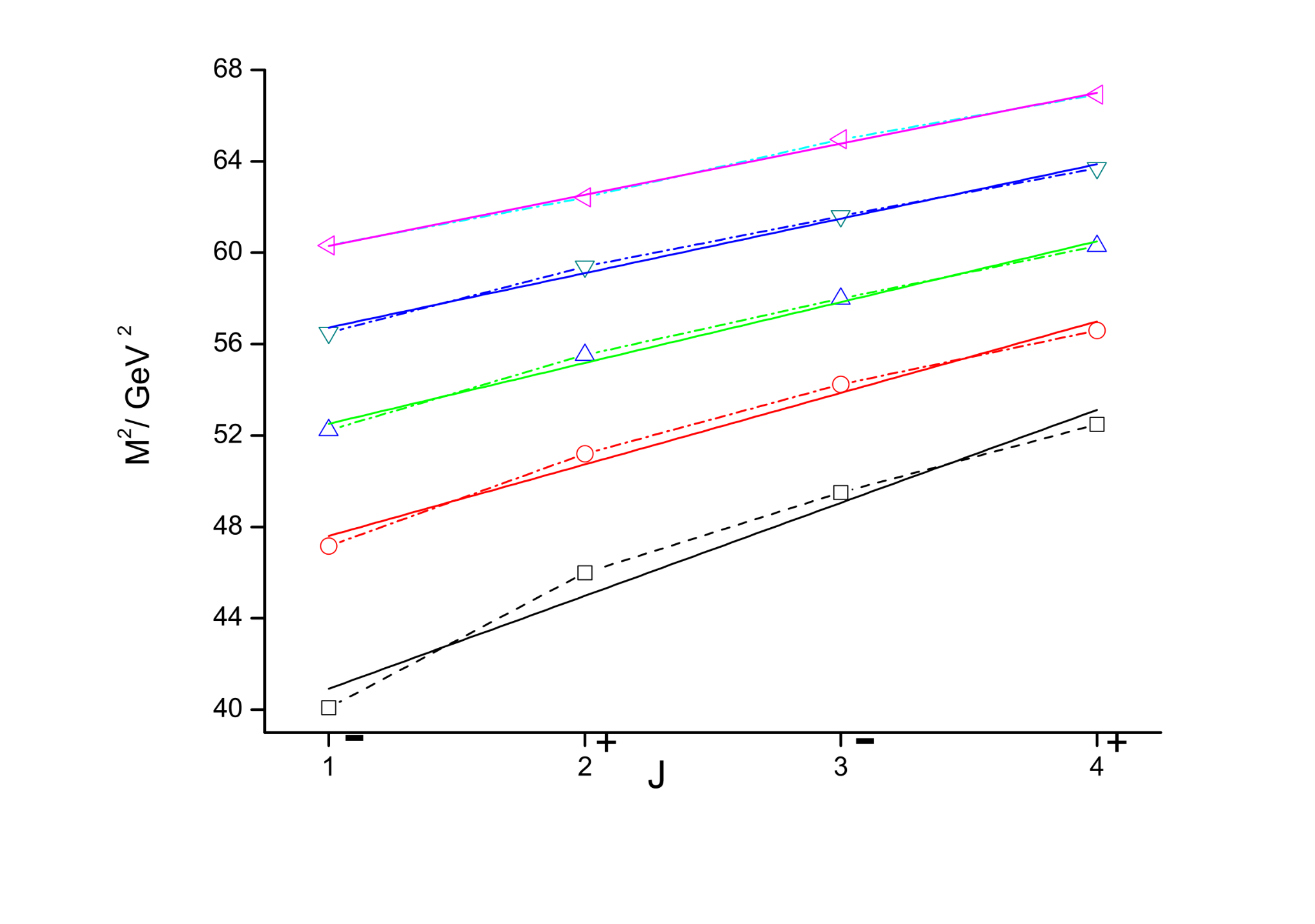}\hfill
	\includegraphics[width=0.48\textwidth]{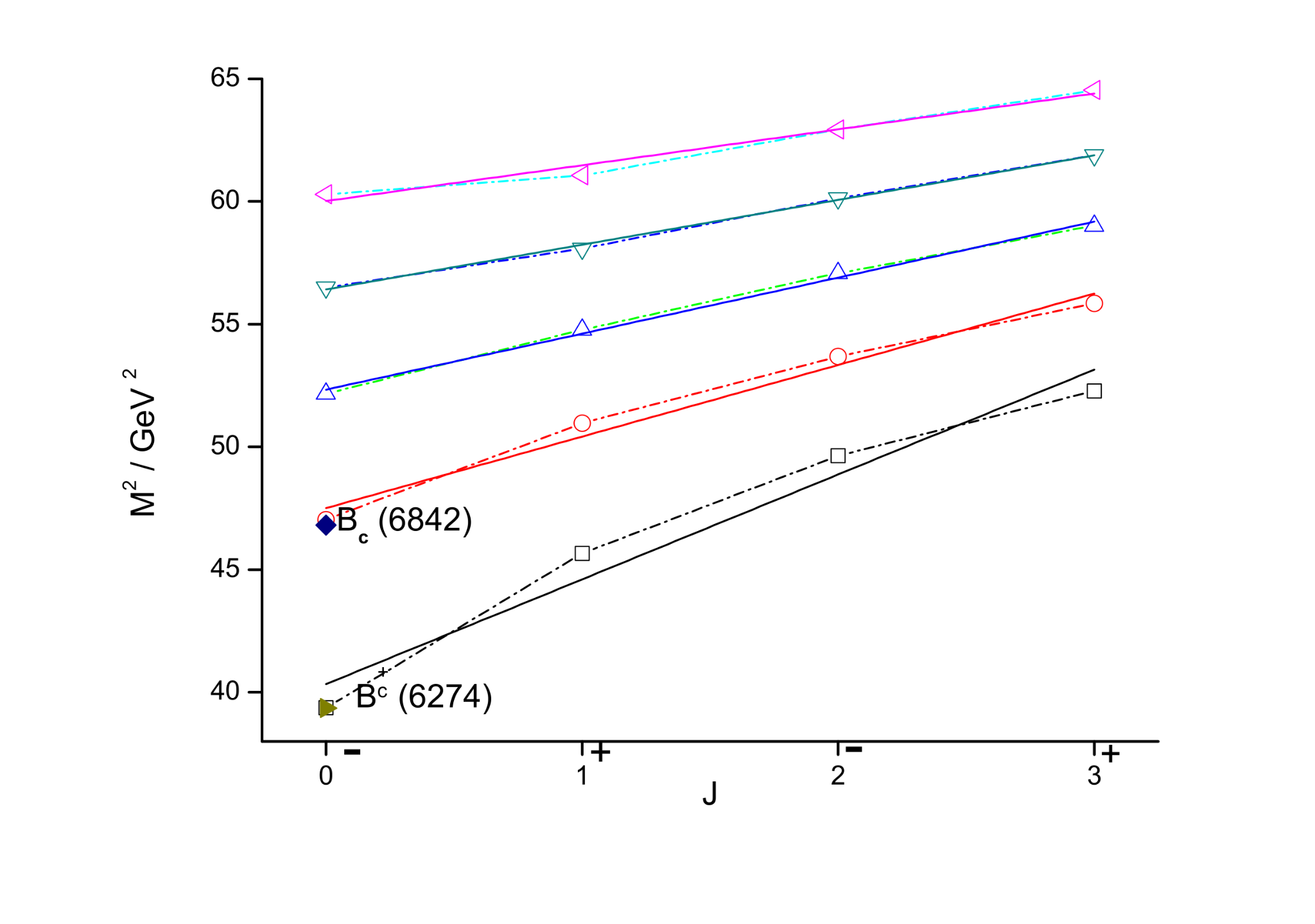}
	\caption{Regge trajectories of the \(B_c\) meson in the \((J,M^2)\) plane: natural-parity states (left) and unnatural-parity states (right). Open symbols denote the model predictions and filled symbols denote experimentally observed states included in the construction.}
	\label{fig:bc_regge_natural}
\end{figure*}

\begin{figure*}[!htb]
	\centering
	\includegraphics[width=0.48\textwidth]{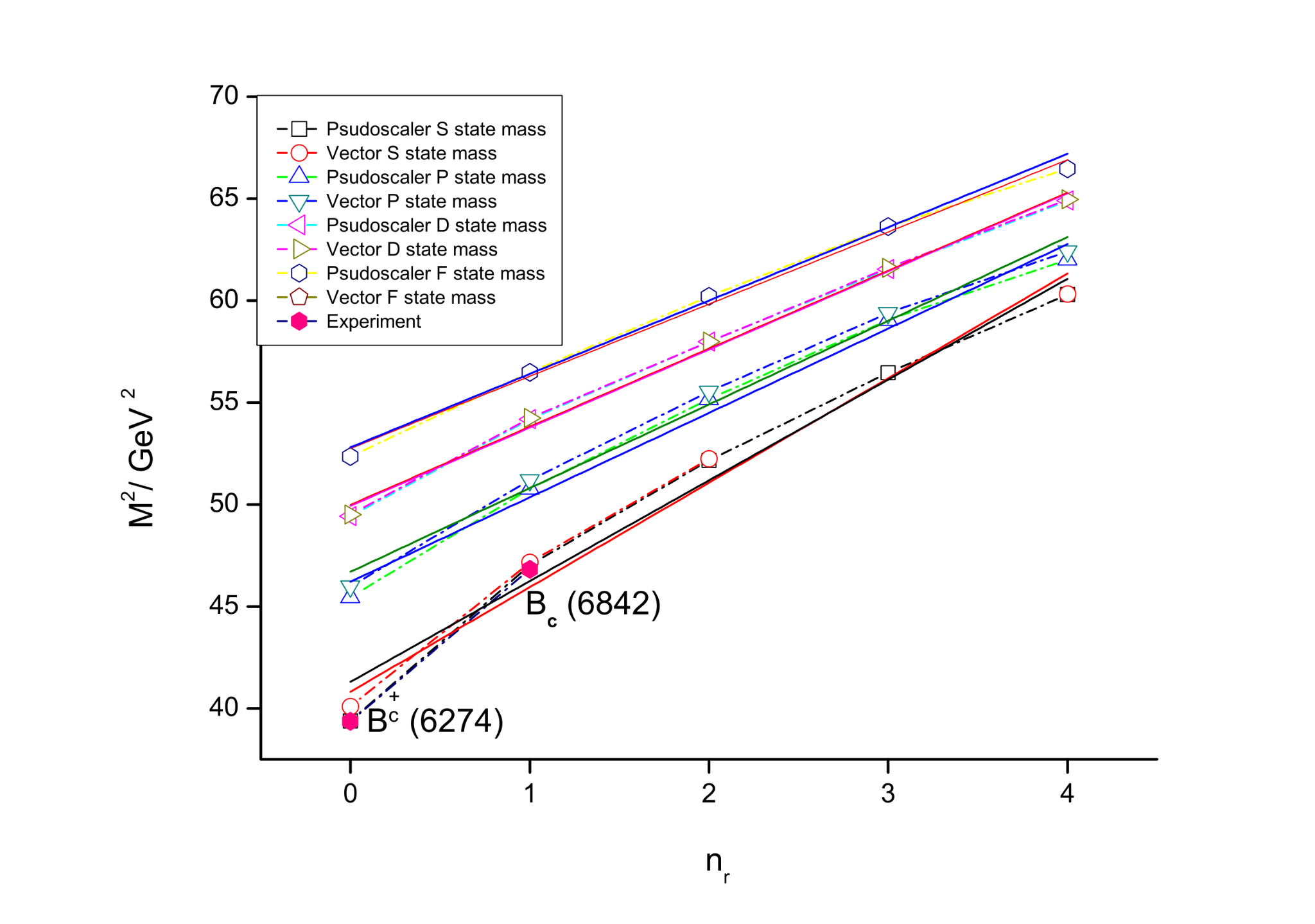}\hfill
	\includegraphics[width=0.48\textwidth]{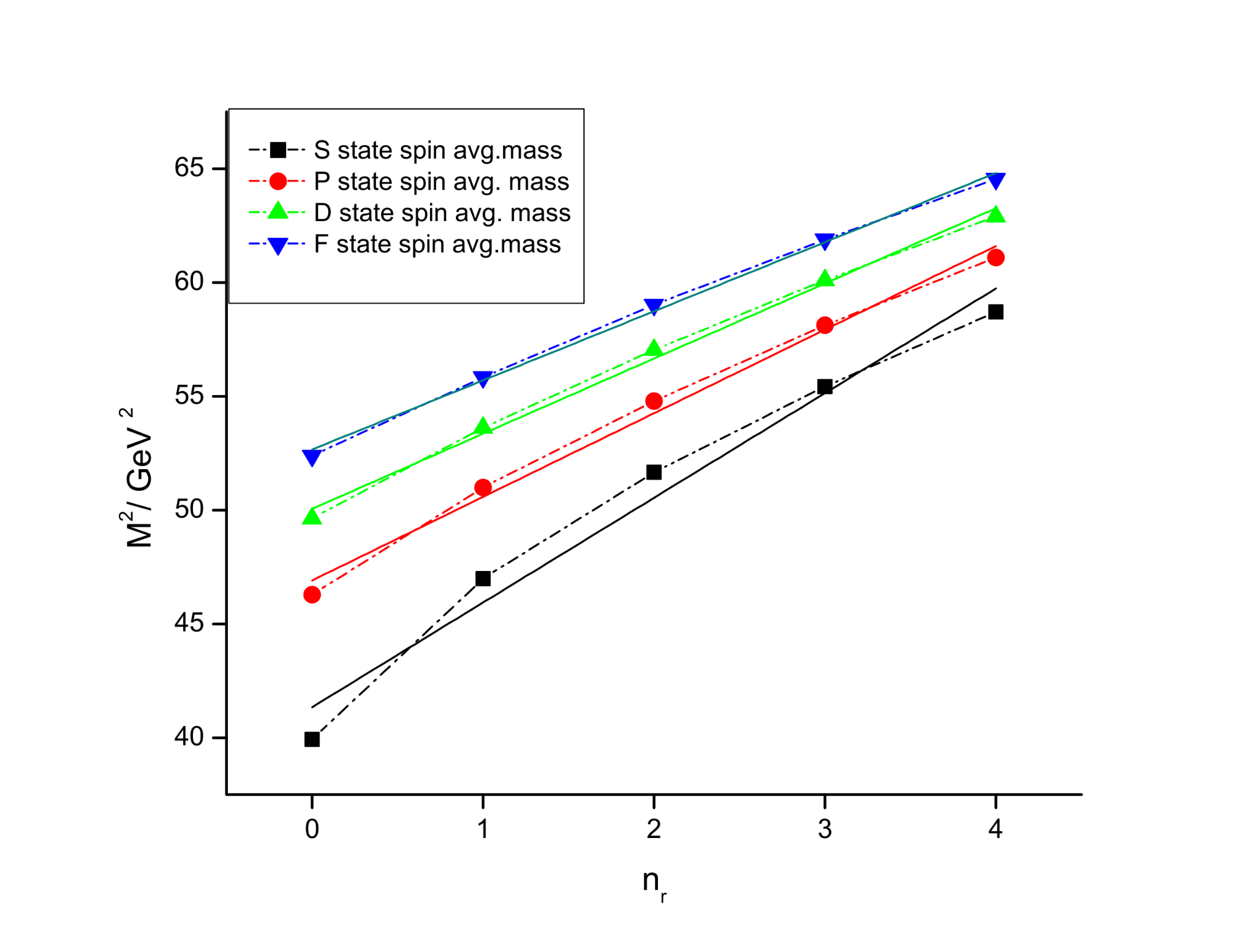}
	\caption{Regge trajectories of the \(B_c\) meson in the \((n_r,M^2)\) plane. The left panel shows trajectories for representative pseudoscalar, vector, and orbitally excited states, while the right panel displays the spin-averaged \(S\)-, \(P\)-, \(D\)-, and \(F\)-wave trajectories.}
	\label{fig:bc_regge_spinavg}
\end{figure*}

\begin{table}[H]
	\caption{Slopes and intercepts of the \((J,M^{2})\) parent and daughter Regge trajectories for the \(B_{c}\) meson.}
	\label{Table:slope1}
	\centering
	\begin{tabular}{cccc}
		\noalign{\smallskip}\hline\noalign{\smallskip}
		\noalign{\smallskip}\hline\noalign{\smallskip}
		Parity & Trajectory & $\alpha\,(\mathrm{GeV}^{-2})$ & $\alpha_{0}$\\
		\noalign{\smallskip}\hline\noalign{\smallskip}
		Natural & Parent & $0.242\pm0.036$ & $-8.941\pm1.172$\\
		Natural & $1^{\mathrm{st}}$ daughter & $0.280\pm0.022$ & $-11.303\pm1.025$\\
		Natural & $2^{\mathrm{nd}}$ daughter & $0.408\pm0.031$ & $-20.201\pm1.741$\\
		Natural & $3^{\mathrm{rd}}$ daughter & $0.464\pm0.036$ & $-24.876\pm2.122$\\
		Natural & $4^{\mathrm{th}}$ daughter & $0.464\pm0.036$ & $-24.876\pm2.121$\\
		\noalign{\smallskip}\hline\noalign{\smallskip}
		Unnatural & Parent & $0.207\pm0.023$ & $-8.30\pm1.132$\\
		Unnatural & $1^{\mathrm{st}}$ daughter & $0.290\pm0.022$ & $-13.703\pm1.185$\\
		Unnatural & $2^{\mathrm{nd}}$ daughter & $0.368\pm0.030$ & $-19.359\pm1.72$\\
		Unnatural & $3^{\mathrm{rd}}$ daughter & $0.424\pm0.029$ & $-23.986\pm1.722$\\
		Unnatural & $4^{\mathrm{th}}$ daughter & $0.469\pm0.030$ & $-28.077\pm1.915$\\
		\noalign{\smallskip}\hline\noalign{\smallskip}
	\end{tabular}
\end{table}

\begin{table}[H]
	\caption{Slopes and intercepts of the \((n_r,M^{2})\) Regge trajectories for representative \(B_c\) states.}
	\label{Table:slope2}
	\centering
	\begin{tabular}{cccc}
		\noalign{\smallskip}\hline\noalign{\smallskip}
		\noalign{\smallskip}\hline\noalign{\smallskip}
		State & $J^P$ & $\beta\,(\mathrm{GeV}^{-2})$ & $\beta_{0}$\\
		\noalign{\smallskip}\hline\noalign{\smallskip}
		$B_{c}^0(6.275)$ & $0^{-}$ & $0.195\pm0.018$ & $-7.950\pm0.973$\\
		$B_{c}^{*}(6.842)$ & $1^{-}$ & $0.202\pm0.017$ & $-8.328\pm0.901$\\
		$B_{c}^{0}$ & $0^{+}$ & $0.237\pm0.016$ & $-10.944\pm0.876$\\
		$B_{c1}^{*}$ & $1^{+}$ & $0.238\pm0.016$ & $-11.112\pm0.899$\\
		$B_{c}^{1}$ & $1^{+}$ & $0.293\pm0.012$ & $-14.983\pm0.718$\\
		$B_{c}^{2}$ & $2^{+}$ & $0.294\pm0.012$ & $-15.024\pm0.707$\\
		$B_{c}$ & $1^{-}$ & $0.319\pm0.010$ & $-17.176\pm0.653$\\
		$B_{c}$ & $2^{-}$ & $0.318\pm0.010$ & $-17.124\pm0.653$\\
		\noalign{\smallskip}\hline\noalign{\smallskip}
	\end{tabular}
\end{table}

\begin{table}[H]
	\caption{Slopes and intercepts of the \((n_r,M^{2})\) spin-averaged Regge trajectories for the \(B_c\) meson.}
	\label{Table:slope3}
	\centering
	\begin{tabular}{ccc}
		\noalign{\smallskip}\hline\noalign{\smallskip}
		\noalign{\smallskip}\hline\noalign{\smallskip}
		Trajectory & $\beta\,(\mathrm{GeV}^{-2})$ & $\beta_{0}$\\
		\noalign{\smallskip}\hline\noalign{\smallskip}
		S State & $0.200\pm0.017$ & $-8.222\pm0.915$\\
		P State & $0.238\pm0.016$ & $-11.075\pm0.893$\\
		D State & $0.293\pm0.012$ & $-14.986\pm0.711$\\
		F State & $0.318\pm0.010$ & $-17.124\pm0.653$\\
		\noalign{\smallskip}\hline\noalign{\smallskip}
	\end{tabular}
\end{table}

The \((J,M^2)\) trajectories shown in Fig.~\ref{fig:bc_regge_natural} are approximately linear within each family and exhibit the familiar separation into parent and daughter trajectories. This is the expected behavior for a well-organized heavy-meson spectrum and agrees with earlier Regge studies of the \(B_c\) system \cite{Ebert2011Regge,DevlaniRai2014,Monteiro2017}. At the same time, the lowest states show mild curvature, especially in the parent trajectories. Such deviations from exact linearity near the ground state are physically natural, because the lowest \(B_c\) levels are more sensitive to the short-distance Coulombic part of the interaction, whereas the higher excitations are governed more strongly by the long-distance confining dynamics and therefore align more smoothly.

The fitted \((J,M^2)\) slopes listed in Table~\ref{Table:slope1} follow the same pattern. For both natural- and unnatural-parity families, the parent trajectories have the smallest slopes, while the daughter trajectories become progressively steeper. This indicates that the level spacing is not strictly uniform across the spectrum and that the softened long-distance dynamics affects the higher families more strongly than the lowest ones. The difference between natural- and unnatural-parity parents is modest, which suggests that the gross organization of the spectrum is not significantly distorted by parity-dependent artifacts.

The \((n_r,M^2)\) trajectories displayed in Fig.~\ref{fig:bc_regge_spinavg} provide a complementary view. These trajectories are also close to linear, both for individual channels and for the spin-averaged orbital families. The extracted slopes in Tables~\ref{Table:slope2} and \ref{Table:slope3} increase systematically from the \(S\)-wave to the \(F\)-wave sectors. This monotonic behavior reflects a progressive change in radial spacing with increasing orbital excitation and is consistent with the general pattern found in earlier Regge analyses of heavy-meson spectra \cite{Ebert2011Regge,DevlaniRai2014,Monteiro2017}. The same trend is visible at the level of individual states: the low-lying \(0^{-}\) and \(1^{-}\) trajectories are the shallowest, while the higher-orbital channels show larger slopes.

An important aspect of the Regge construction is that it summarizes, in a compact visual form, the same information already visible in the mass tables. The detailed spectroscopy indicates that the screened-potential framework reproduces the accepted low-lying structure while yielding a somewhat softer upper spectrum than many unscreened models. The Regge plots express this same result globally: the higher trajectories are close to linear and therefore confirm the coherence of the excited-state spectrum, while the residual curvature of the parent lines reflects the persistence of short-distance effects in the lowest states. The Regge analysis therefore does not introduce a separate phenomenology so much as it confirms, in a condensed form, the same spectral regularity that underlies the weak and radiative sectors.

From a phenomenological viewpoint, this subsection is also timely because several excited \(B_c\) structures have now been observed experimentally. Although present data are still insufficient to determine full trajectory parameters directly from experiment, the existence of \(2S\)- and \(1P\)-wave signals means that the global organization of the \(B_c\) family is no longer purely speculative. The Regge analysis therefore provides a useful bridge between the detailed screened-potential predictions and the gradually expanding experimental map of excited \(B_c\) states.

\subsection{Overall phenomenological picture}

Taken together, the present study yields a coherent phenomenological description of the \(B_c\)-meson system. At the spectroscopic level, the low-lying states remain close to the established experimental and quark-model benchmarks, while the higher excitations display the moderate compression expected in a screened-potential framework. At the dynamical level, the same wave functions generate decay constants of the correct order, weak transition amplitudes with clear channel dependence, radiative widths that reproduce the standard hierarchy between dominant \(E1\) and suppressed \(M1\) transitions, and Regge trajectories that organize the spectrum into approximately linear families. The resulting picture is therefore not fragmented: the spectroscopic, weak, radiative, and global trajectory sectors all point to the same underlying dynamics.

A particularly valuable aspect of the analysis is that the channel dependence seen in one sector reappears in a consistent way elsewhere. The enhanced sensitivity of the \(\eta_c\) modes in the \(S\)-wave weak sector, the dominance of \(h_c\) among the \(P\)-wave channels, the stable but strongly suppressed \(D\)-wave hierarchy, and the orderly radiative cascade pattern are not isolated numerical features, but related consequences of the same spectral and wave-function input. In this sense, the main strength of the work is not one single numerical agreement, but the ability of one common framework to describe a wide range of observables without losing internal coherence.

At the same time, the analysis also identifies the sectors in which the framework remains incomplete. The inclusive lifetime is longer than the current experimental benchmark, and selected nonleptonic normalizations indicate that some channels remain more model-sensitive than others. These tensions should be viewed as useful diagnostics rather than as failures of principle, because they isolate the observables for which improvements in short-distance coefficients, decay constants, form factors, factorization, or inclusive decay mechanisms are most needed. The overall conclusion is therefore twofold: the screened-potential framework already provides a broad and internally consistent description of the \(B_c\) system across spectroscopy and decay phenomenology, and the remaining discrepancies define a clear direction for future refinement.

\section{Conclusion}

In this work, the spectroscopy and decay properties of the \(B_c\) meson have been investigated in a screened-potential framework with spin-dependent interactions and a unified treatment of the associated phenomenology. The main objective was to test whether a single bound-state description can reproduce the known low-lying spectrum, provide plausible predictions for higher excitations, and yield a coherent pattern across weak, radiative, and global spectroscopic observables.

The calculated low-lying \(B_c\) states remain close to the established experimental and theoretical benchmarks, while the higher \(S\)-, \(P\)-, \(D\)-, and \(F\)-wave excitations show the moderate compression expected from screened confinement. The resolved spectrum preserves the standard fine and hyperfine ordering, indicating that the screened interaction modifies the upper part of the spectrum without disturbing the accepted short-distance structure. The decay constants support the same picture: the ground-state values remain within the range of modern theoretical estimates, whereas the stronger suppression of higher radial excitations emerges as a characteristic consequence of screening.

The weak-decay analysis reveals a stable hierarchy across all orbital sectors. In the \(S\)-wave sector, the corrected semileptonic observables, including \(R(\eta_c)\), \(R(J/\psi)\), and the polarization quantities, are in good agreement with recent Standard Model benchmarks, and several experimentally relevant branching-fraction ratios are reproduced at a reasonable level. In the \(P\)-wave sector, the dominance of the \(h_c\) channels appears as a robust structural result, although some nonleptonic normalizations remain more model-dependent. In the \(D\)-wave sector, the predicted suppression pattern is stable and provides a useful set of forecasts for future measurements. The analysis of nonleptonic channels involving charmonium together with \(D^{(*)}_{(s)}\) mesons further confirms the expected flavor hierarchy, especially the enhancement of strange final states, although the normalization of the \(J/\psi D_s^{(*)}\) modes still requires improvement.

The purely leptonic, annihilation, radiative, and Regge sectors provide additional consistency checks. The leptonic and annihilation observables indicate that the short-distance normalization is of the correct order, but the predicted lifetime remains longer than the experimental benchmark, showing that the total weak width is still underestimated. The radiative widths reproduce the standard hierarchy of dominant \(E1\) and suppressed \(M1\) transitions, while the Regge trajectories confirm that the calculated spectrum is globally well organized and approximately linear in both the \((J,M^2)\) and \((n_r,M^2)\) planes.

Overall, the present study provides a unified phenomenological description of the \(B_c\) meson in which spectroscopy, decay constants, weak transitions, radiative processes, and Regge behavior are described within one common framework. Its main strength lies in this internal consistency across different observables. At the same time, the remaining tensions, particularly in the inclusive lifetime and selected nonleptonic normalizations, indicate where further refinement is needed. Future improvements in the treatment of short-distance coefficients, nonfactorizable effects, coupled-channel dynamics, and inclusive decay mechanisms should make this framework more competitive for precision studies of the \(B_c\) system.

\end{document}